\documentclass[a4paper,fleqn,usenatbib]{mnras}
\usepackage{newtxtext,newtxmath}
\usepackage[T1]{fontenc}
\usepackage{ae,aecompl}

\usepackage{graphicx}	
\usepackage{amsmath}	
\usepackage{amssymb}	
\usepackage{mathrsfs}
\usepackage{listings}
\usepackage{etoolbox}
\usepackage{cuted}
\usepackage{flushend}
\usepackage{breqn}
\usepackage{todonotes}
\usepackage{newtxtext,newtxmath}
\usepackage{hyperref}
\makeatletter
\preto\gather{\@fleqnfalse}
\makeatother

\DeclareSymbolFont{CMletters}{OML}{cmm}{m}{it}
\DeclareMathSymbol{v}{\mathord}{CMletters}{`v}
\DeclareMathOperator{\sinc}{sinc}
\DeclareFontFamily{U}{wncy}{}
\DeclareFontShape{U}{wncy}{m}{n}{<->wncyr10}{}
\DeclareSymbolFont{mcy}{U}{wncy}{m}{n}
\DeclareMathSymbol{\Sh}{\mathord}{mcy}{"58}

\title[Optimal gridding and degridding]{Optimal gridding and degridding in radio interferometry imaging}

\author[H. Ye et al.]{
Haoyang Ye,$^{1}$
Stephen F. Gull,$^{1}$\thanks{E-mail: sfg1@cam.ac.uk}
Sze M. Tan $^{2}$
Bojan Nikolic$^{1}$
\\
$^{1}$Astrophysics Group, Cavendish Laboratory, University of Cambridge, Cambridge CB3 0HE, UK\\
$^{2}$Picarro, Inc., 3105 Patrick Henry Dr., Santa Clara, CA 95054, USA
}

\date{Accepted 14 Oct 2019. Received 16 June 2019; in original form 14 Oct 2019}

\pubyear{2019}

\begin{document}
\label{firstpage}
\pagerange{\pageref{firstpage}--\pageref{lastpage}}
\maketitle

\begin{abstract}
In radio interferometry imaging, the Fast Fourier transform (FFT) is often used to compute maps from visibility data. A gridding procedure for convolving the measured visibilites with a chosen gridding function is used to transform visibility values into uniformly sampled grid points. We propose here a parameterised family of `least-misfit gridding functions' which minimise an upper bound on the difference between the DFT and FFT dirty images for a given gridding support width and image cropping ratio. When compared with the widely used spheroidal function with similar parameters, these provide more than $100$ times better alias suppression and RMS misfit reduction over the usable dirty map. We discuss how appropriate parameter selection and tabulation of these functions allow for a balance between accuracy, computational cost and storage size. Although it is possible to reduce the errors introduced in the gridding or degridding process to the level of machine precision, accuracy comparable to that achieved by \texttt{CASA} requires only a lookup table with $300$ entries and a support width of $3$, allowing for a greatly reduced computation cost for a given performance. \end{abstract}




\section{Introduction}
In radio interferometry, the relation between the visibility data $V$ and the sky brightness distribution $I$ is derived with clarity by \citet{1999ASPC..180....1C} and \citet{thompson1999fundamentals}. Using the $(u,v,w)$ and $(l,m,n)$ coordinate systems defined in \citet{thompson1999fundamentals}, this relation can be expressed as
\begin{multline}\label{eq:fourier_visibility}
	V(u,v,w) = \int\int \frac{\text{d}l \text{d}m }{\sqrt{1-l^2-m^2}}\\
	I(l,m)\exp\left[-i2\pi\left(ul+vm+w\left(\sqrt{1-l^2-m^2}-1\right)\right)\right],	
\end{multline}
where $(l,m)$ are direction cosines between $-1$ and $1$, and $(u,v,w)$ are baseline coordinates in units of wavelength.

If the field of observation is very small, and is close to the phase centre, then $w(\sqrt{1-l^2-m^2}-1) \approx 0$ can be neglected. The visibility function can then be written as
\begin{equation}\label{eq:inversetransform}
	V(u,v)= \int \int \text{d}l \text{d}m I(l,m) \exp[-i2\pi(ul+vm)].
\end{equation}
Thus, the sky brightness can be obtained by performing an inverse two--dimensional Fourier transform on the visibility function. Owing to the incomplete sampling of the $(u,v)$ plane, the result of taking the inverse direct Fourier transform (DFT) of the sampled and weighted visibility data $V$ is a `dirty image', rather than the true sky brightness. The inverse DFT of the sampling function is referred to as the synthesised beam, or the dirty beam.

When the number of visibilities $N_{v} \gg \log (N_xN_y)$, computation of the DFT becomes computationally expensive relative to using the FFT, which reduces the computational complexity of making an image of size $N_x$ pixels by $N_y$ pixels from $\mathcal{O}(N_{v}N_{x}N_{y})$ to $\mathcal{O}\bigg (N_{x}N_{y} \log (N_{x}N_{y})\bigg ) + \mathcal{O}(N_{v})$ operations \citep{10.2307/2003354, heideman1985gauss, 2017A&A...603A..40S}.

The FFT algorithm requires the data to be sampled on a Cartesian grid, however, which is not the case for visibility data. The obvious solution is to interpolate the visibility data onto a Cartesian grid and then apply the FFT. The nearest--neighbour method was an early application of gridding in radio astronomy, and an example of its implementation is given by \citet{hogg1969synthesis}. More elaborate interpolation methods include `cell summing' in \citet{mathur1969pseudodynamic}, and `radial interpolation' proposed by \citet{thompson1974interpolation}. These interpolation methods are no longer used, because they proved to be poor at suppressing aliasing. Aliasing is an unwanted phenomenon in which brightness, including noise outside the field of interest, appears within the field of interest \citep{greisen1979effects, 1999ASPC..180..127B}.

The most common method used to overcome this difficulty is to convolve the visibility data with a gridding function. The gridding function $C(u,v,w)$ should satisfy the following three conditions \citep{greisen1979effects}:
\begin{enumerate}
	\item Separability: $C(u,v,w)=C_u(u)C_v(v)C_w(w)$, where $C_u$, $C_v$ and $C_w$ are usually chosen to be the same function.
	\item Be real and symmetric about the origin. The one--dimensional form $C(u)$ is used for simplicity from now on. 
	\item Be nonzero only within a finite window, so that the computational cost of the convolution is not unduly large.
\end{enumerate}

The gridding process can be viewed mathematically as a combination of a convolution and a sampling process \citep{o1985fast}. In practice, the convolution is conducted only at each grid point \citep{thompson1986interferometry}. An image cropping process is usually performed, since the errors at the edge are several orders of magnitude worse than around the centre of the image. As a result, a larger dirty image is always made. Multiplication with a correcting function is then needed to cancel the effects arising from $C(u)$, so as to obtain the correct fluxes. \citet{greisen1976effects} proposed that the correcting function should be the reciprocal of the inverse Fourier transform of $C(u)$. We thus obtain the FFT dirty image, which is an approximation to the DFT dirty image. 

 As early as 1961, Elizabeth Waldram, in the Radio Astronomy Group at the Cavendish Laboratory, Cambridge, was using gridding functions that included the Gaussian, the $\sinc$ function, and a Gaussian times $\sinc$ function, according to her work notes, although this work was unpublished. \citet{1971PhDT.......153B} later used a Gaussian gridding function with the east--west synthesis array at Westerbork. Many gridding functions have since been explored.

To assist in suppressing the effects of aliasing in the dirty image, \citet{1975MComP..14R.131B} proposed a measure of the suppression. It was claimed that the prolate spheroidal wave function of order 0 \citep{6773659, 6773660} should be the optimal gridding function \citep{1975MComP..14R.131B, 1984iimp.conf..333S}. \citet{schwab1980optimal, 1984iimp.conf..333S} then argued for a modified optimality criterion involving weights, which singled out the spheroidal function \citep{1935PNAS...21...51S} as the best gridding function. The spheroidal function is widely implemented in imaging pipelines such as \texttt{AIPS} (Astronomical Image Processing Software \citep{1985daa..conf..195W}) and \texttt{CASA} (Common Astronomy Software Applications \citep{2007ASPC..376..127M}). The spheroidal function \footnote{$\alpha = 1$ is chosen, since it suppresses aliasing more effectively than $\alpha = 0$.} will be the benchmark for comparison in this paper. In our numerical experiments, we use the `\texttt{pro\_ang1}' function from the \texttt{Python} package \texttt{scipy}. The root mean square (RMS) difference between this function and the numerical approximation proposed by \citet{schwab1981vla} is less than $10^{-6}$.

Use of the gridding function is not confined to the making of dirty images and dirty beams, for it is also used in the degridding process. Gridding and degridding are mathematically transpose operations; degridding reconstructs the visibility data from a given image model. It is essential to deconvolution methods such as Cotton--Schwab CLEAN \citep{1983AJ.....88..688S} and the Maximum Entropy method \citep{1978Natur.272..686G}. Additionally, degridding is also used during the self--calibration procedure \citep{1999ASPC..180..187C}. In summary, the gridding function is required in both gridding and degridding in imaging procedures, and its choice consequently influences the quality of the images obtained, and all analyses based on them.



A systematic criterion should be sought for selecting the gridding function in view of its major role in the imaging and self--calibration processes. The gridding function was originally introduced to further the replacement of DFT by FFT, and any new gridding function should therefore minimise the difference between the DFT and FFT dirty images. By finding and implementing such a gridding function, we are able to obtain results which better approximate the ideal DFT results. Image--based data analysis, such as source extraction from dirty images as explained in \citet{hague2018bayesian}, should also benefit from improved image accuracy. We will also be able to obtain more accurate degridded/self--calibrated visibilities.

This paper develops the subject of gridding functions, and also the processes of gridding and degridding. In Section~\ref{sec:c4s2} a new gridding function, the `least--misfit gridding function\footnote{A series of Jupyter notebook tutorials describing the use and properties of the least--misfit functions can be found at \url{https://github.com/SzeMengTan/OptimalGridding}}', is proposed, based on minimising the upper bound of the difference between the DFT and FFT dirty images. The theory, and its computational implementation, are explained in full detail. The spheroidal function is reevaluated according to the same criterion. In Section~\ref{sec:c4s4}, the use of the spheroidal function and the least--misfit function are compared theoretically and numerically by examining the resulting image accuracy and suppression of aliasing. Section~\ref{sec:degridding} shows the degridding performance using the least--misfit gridding function. Practical implementation of the least--misfit functions in imaging is demonstrated in Section~\ref{c4sec:computaional_cost}, including calculation of the computational cost.

\section{Least--misfit gridding function}\label{sec:c4s2}

The inverse relationship to Equation (\ref{eq:inversetransform}) for recovering the sky brightness from the visibility plane data is
	\[ I(l,m) = \iint \mathrm{d}u\,\mathrm{d}v\, V(u,v) \exp[i2\pi(ul + vm)] \]
	When mapping a small portion of the sky, say $-L/2\leq l\leq L/2$ and $-M/2\leq m\leq M/2$ it is convenient to introduce normalised map coordinates $(x,y)$ defined by $x=l/L, y=m/M$ which each range from $-\frac{1}{2}$ to $\frac{1}{2}$. Writing $u'=uL$ and $v'=vM$, we find
	\[ I(x,y) \propto  \iint \mathcal{V}(u',v')  \exp\left[\mathrm{i}2\pi\left(u' x + v' y\right)\right] \,\mathrm{d}u'\,\mathrm{d}v'\]
	In the following, we drop the primes on $u'$ and $v'$ for convenience, so they are now in units of cell widths rather than wavelengths, and use them in conjunction with the normalised map coordinates $x$ and $y$.

\begin{figure}
\centering\includegraphics[width=0.8\columnwidth]{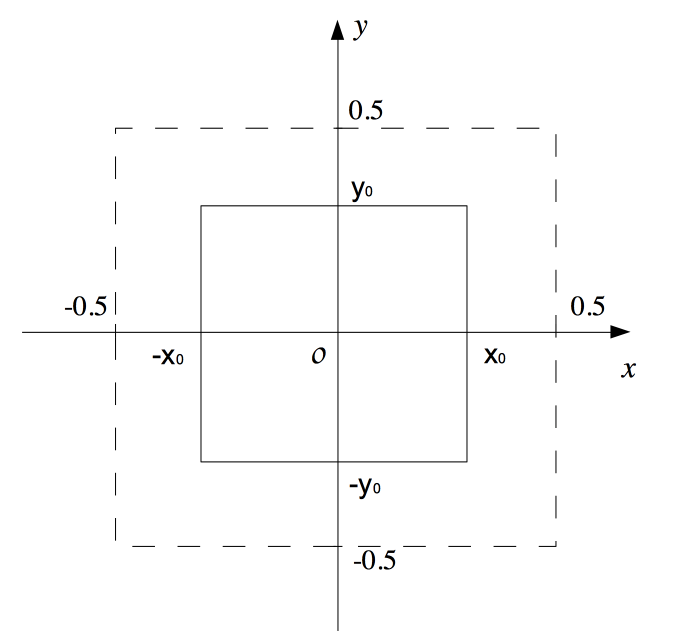}
\caption{The normalised coordinate system $(x,y)$. The size of the two--dimensional field of interest is defined by $x\in [-x_0,x_0]$ and $y\in [-y_0,y_0]$.}\label{fig:x0_coordinate}
\end{figure}

With infinite computing power, DFT would be preferable to FFT, since no information would be lost due to gridding or FFT. We therefore propose a criterion which minimises the upper bound of the difference between the DFT and FFT dirty images. This difference is also referred to below as the `image misfit'. Since cropping of the edge of the image is common in order to maintain a reasonably good image misfit across the retained image, we introduce a parameter $x_0$ to control the retained portion of the image, so as to minimise the image misfit within the field of view of interest. Figure (\ref{fig:x0_coordinate}) shows the normalised coordinate system, where the field of view of interest is $x\in [-x_0,x_0]$, $y\in [-y_0,y_0]$ where $x_0, y_0 \in (0,0.5]$. The parameters $x_0$ and $y_0$ control the amount of discard in both directions. They are usually taken to be the same, but that does not mean the angular sizes of both sides of the field are identical. \citet{o1985fast} proposed that an FFT image should be made twice as large as the intended image, so that the outer half of it should be discarded; in this case $x_0 = y_0 = 0.25$.

Our aim is to find a gridding function which can minimise the upper bound of the dirty image misfit within the desired portion of the image.

\subsection{Theory}

We first write the DFT and FFT dirty image so as to find an expression for the image misfit. We perform the analysis in one dimension for simplicity. The DFT dirty image can be written as
\begin{equation}\label{eq:dirty_image_1D}
	I_D(x) = \sum_{k} w_k V_k \exp{(i2\pi u_k x)},
\end{equation}
where $w_k$ is the weight for the visibility data $V_k$.

To obtain the corresponding FFT dirty image $\widehat{I}_D(x)$ with size $N_x$, the visibilities must be gridded to obtain the gridded data $G_n$, after which the FFT is applied
\begin{equation}
	\widehat{I}_D(x) = h(x)\sum^{N_x/2-1}_{n=-N_x/2}G_n \exp{(i2 \pi  nx)},
\end{equation}
where $h(x)$ is the correcting function. The coefficient $G_n$ is written as
\begin{equation}
	G_n = \sum_{k} w_k V_k C(n-u_k).
\end{equation}
The gridding function $C(u)$ is real and symmetric, with a support of $W$ cells of unit width. Each visibility $V_k$ is assigned with differing weights obtained from $C(u)$ onto $W$ consecutive grid points $n \in S_k$, where
\begin{equation}
	S_k = \{n \in \mathbb{Z}: u_k-W/2\leq n<u_k+W/2\}.
\end{equation}
The integer $W$ is chosen by the user.

Minimisation of the misfit between the DFT and FFT dirty image was first proposed and implemented by \citet{tan1986aperture}. This work is extended here, with new results. We begin by writing the error resulting from the use of $\widehat{I}_D$ instead of $I_D$ as
\begin{align}\label{eq:ex}
	e(x) \equiv I_D(x) - & \widehat{I}_D(x)  =\sum_{k} w_k\underbrace{V_k \exp{(i2\pi u_k x)}}_{a_k^*}\nonumber \\
	& \underbrace{\left (1-h(x)\sum_{n\in S_k} C(n-u_k)\exp{[i2\pi (n-u_k)x]}\right )}_{b_k}.
\end{align}
According to the Cauchy--Schwarz inequality
\begin{equation}
	|\textbf{a}\cdot \textbf{b}|^2 \leq |\textbf{a} \cdot \textbf{a}| |\textbf{b} \cdot \textbf{b}|, \text{ with } \textbf{a} \cdot \textbf{b}\equiv \sum w_k a_k^*b_k,\nonumber
\end{equation}
the following expression can be derived:  
\begin{strip}
\begin{gather}\label{eq:e(x)2}
	|e(x)|^2\leq \bigg (\sum_{k} w_k |V_k|^2 \bigg ) \Bigg(\underbrace{\sum_{k} w_k \bigg |1-h(x)\sum_{n\in S_k} C(n-u_k)\exp{[i2\pi (n-u_k)x]\bigg |^2}}_{\ell(x)}\Bigg ).
\end{gather}
\hrule
\end{strip}

Since the term $\sum  w_k |V_k|^2$ depends only on the data and the weights, our task is reduced to keeping the other factor $\ell(x)$ small by choosing $C(u)$ and $h(x)$ appropriately.

By writing the fractional offset part of $u$ as $\nu = u - \lfloor u \rfloor\in [0,1]$, where $\lfloor u \rfloor$ is the largest integer not greater than $u$, we have
\begin{equation}
	\sum_{n\in S_k}C(n-u_k)\exp[i2\pi(n-u_k)x] \equiv \sum_r C(r-\nu_k)\exp[i2\pi (r-\nu_k)x],
\end{equation}
where $r=-W/2+1,...,W/2$ when W is even. When $W$ is odd with $u_k-\lfloor u_k \rfloor>0.5$, $r = (3-W)/2,...,(W+1)/2$; if $u_k-\lfloor u_k \rfloor \leq 0.5$, then $r = (1-W)/2,...,(W-1)/2$.

Hence, we can write
\begin{equation}\label{eq:ell}
	\ell(x) = \sum_{k} w_k \bigg |1-h(x)\sum_r C(r-\nu_k)\exp{[i2\pi (r-\nu_k)x]\bigg |^2}.
\end{equation}

When the number of visibilities $M$ is large, the values of $\nu$ are effectively spread randomly throughout the interval $(0,1)$. The sum inside Equation (\ref{eq:ell}) can therefore be replaced accurately by an integral from 0 to 1 over $\nu$, provided that the weights are normalised to $\sum_k w_k = 1$. Hence, we have the general dimensionless local error limit
\begin{align}\label{eq:l(x)}
\ell(x) &= \int_0^1\textrm{d}\nu \bigg| 1-h(x)\sum_r C(r-\nu)\exp[i2\pi (r-\nu)x]\bigg|^2 \\
 		& = 1-2h(x)\int_0^1\textrm{d}\nu\sum_r C(r-\nu)\cos[2\pi(r-\nu)x] \nonumber \\
		& +h(x)^2\int_0^1\textrm{d}\nu\sum_{r_1}\sum_{r_2} C(r_1-\nu)C(r_2-\nu)\cos[2\pi(r_1-r_2)x].\nonumber
\end{align}
In this way we can minimise the upper bound of $e^2(x)$ by minimising $\ell(x)$, which we define as the `map error function'. Since $\ell(x)$ depends on both the gridding function and the correcting function, the choice of these is crucial in reaching a small value of $e(x)$. We take $\ell(x)$ as a quantitative measure of the performance of a given gridding function, in terms of the upper bound of the difference between the DFT and FFT dirty images. For a given convolution function $C(u)$, we can minimise the value of $\ell(x)$ at each $x$ by choosing $h(x)$ such that $\displaystyle \frac{\partial l(x)}{\partial h(x)} = 0$. This leads to
\begin{equation}\label{eq:h(x)}
	h(x) = \frac{\int_0^1\mathrm{d}\nu\sum_r C(r-\nu)\cos[2\pi(r-\nu)x]}{\int_0^1\mathrm{d}\nu\sum_{r_1}\sum_{r_2}C(r_1-\nu)C(r_2-\nu)\cos[2\pi(r_1-r_2)x]} .
\end{equation}
The numerator is equal to $c(x)$, the real part of the inverse Fourier transform of $C(u)$, since
\begin{equation}\label{eq:ft_griddingfunc}
	c(x) = \int \mathrm{d}uC(u)\cos (2 \pi ux) = \int_0^1 \mathrm{d}\nu \sum_r C(r-\nu)\cos[2\pi (r-\nu)x].
\end{equation}
According to Equation 2.13 of \citet{tan1986aperture}, the denominator can be shown to be equal to
\begin{equation}\label{eq:maybe}
	\sum_{n=-\infty}^{\infty} c(x-n)^2 = \int_0^1\textrm{d}\nu\sum_{r_1}\sum_{r_2} C(r_1-\nu)C(r_2-\nu)\cos[2\pi(r_1-r_2)x],
\end{equation}
so that the optimal correction function can be written as
\begin{equation}\label{eq:correction_func}
	h(x) = c(x)\bigg/ \sum_{n=-\infty}^{\infty} c(x-n)^2.
\end{equation}
The denominator is the sum of aliased copies of $c(x)^2$ shifted by the integer $n$. The Fourier transform of a good gridding function falls to zero rapidly outside of the map, (i.e., outside $|x|<\frac{1}{2}$) and so to a good approximation, the denominator within the interior of map simplifies to $c(x)^2$, making $h(x)\approx 1/c(x)$, which is the form used by \citet{greisen1976effects}.

The function $\ell(x)$ can measure only the upper bound of the image misfit at certain positions. The overall upper bound of the map error can be written as a normalised integral of the map error function $\ell(x)$ across the retained range $-x_0 \leq x \leq x_0$, as follows
\begin{equation}\label{eq:Error_simple}
	E = \frac{1}{2x_0}\int_{-x_0}^{x_0}\ell(x)\mathrm{d}x.
\end{equation}
The integral is considered over a restricted range because the rest of the dirty image will be discarded. We refer to the gridding function $C(u)$ found by minimising $E$ as the \textit{least--misfit gridding function}. 

Although the minimisation of $E$ is over variations of both $h(x)$ and $C(u)$, we may reduce the problem to consider varying either $C(u)$ alone, or $h(x)$ alone. This can either be done by using Equation (\ref{eq:h(x)}) to express $h(x)$ in terms of $C(u)$, or by using the following procedure to express $C(u)$ in terms of $h(x)$. This alternative has been found to be preferable in practice. From Equation (\ref{eq:l(x)}), it is evident that $\ell(x)$ is the integral over $\nu$ of a non-negative function. If $h(x)$ is given, we can minimise $E$ by choosing $C_r(\nu)\equiv C(r-\nu)$ so as to minimise the integrand in Equation (\ref{eq:l(x)}) for each $\nu$. Upon substituting Equation (\ref{eq:l(x)}) into Equation (\ref{eq:Error_simple}) and following the analysis of \citet{tan1986aperture}, we differentiate $E$ with respect to $C_r(\nu)$ to obtain the set of simultaneous equations for $C_r(\nu)$,
\begin{equation}\label{eq:grid_corec_eq}
	\resizebox{0.5\textwidth}{!}{$\sum_r\int_{-x_0}^{x_0} \mathrm{d}x h(x)^2 \cos [2\pi (r'-r)x] C_r({\nu}) = \int_{-x_0}^{x_0}\mathrm{d}x h(x)\cos [2\pi (r'-\nu)x]$}.	
\end{equation}
where the indices $r$ and $r'$ are dummy indices, which play the same roles as $r_1$ and $r_2$ in Equation (\ref{eq:l(x)}) and (\ref{eq:h(x)}). 

The linear system of equations for $C_r(\nu)$ may then be written in matrix form as
\begin{equation}\label{eq:grid_corec_eq2}
	\sum_{r}A_{r'r}C_r(\nu) = B_{r'} \equiv \int_{-x_0}^{x_0}\mathrm{d}xh(x) \cos[2\pi(r'-\nu)x],
\end{equation}
where $\mathbf{A}$ is a Toeplitz matrix whose elements are independent of $\nu$ and can be written as
\begin{equation}\label{eq:All'}
	A_{r'r} \equiv \int_{-x_0}^{x_0}\mathrm{d}xh(x)^2\cos[2\pi(r'-r)x].
\end{equation}
For each $\nu$, the values $C_r(\nu)$ give the function $C(u)$ at $W$ points. The detailed algorithm for numerical optimisation is set out in Appendix \ref{app:1}.

\subsection{Results and discussion}

\begin{figure}
  \centering\includegraphics[width=\columnwidth]{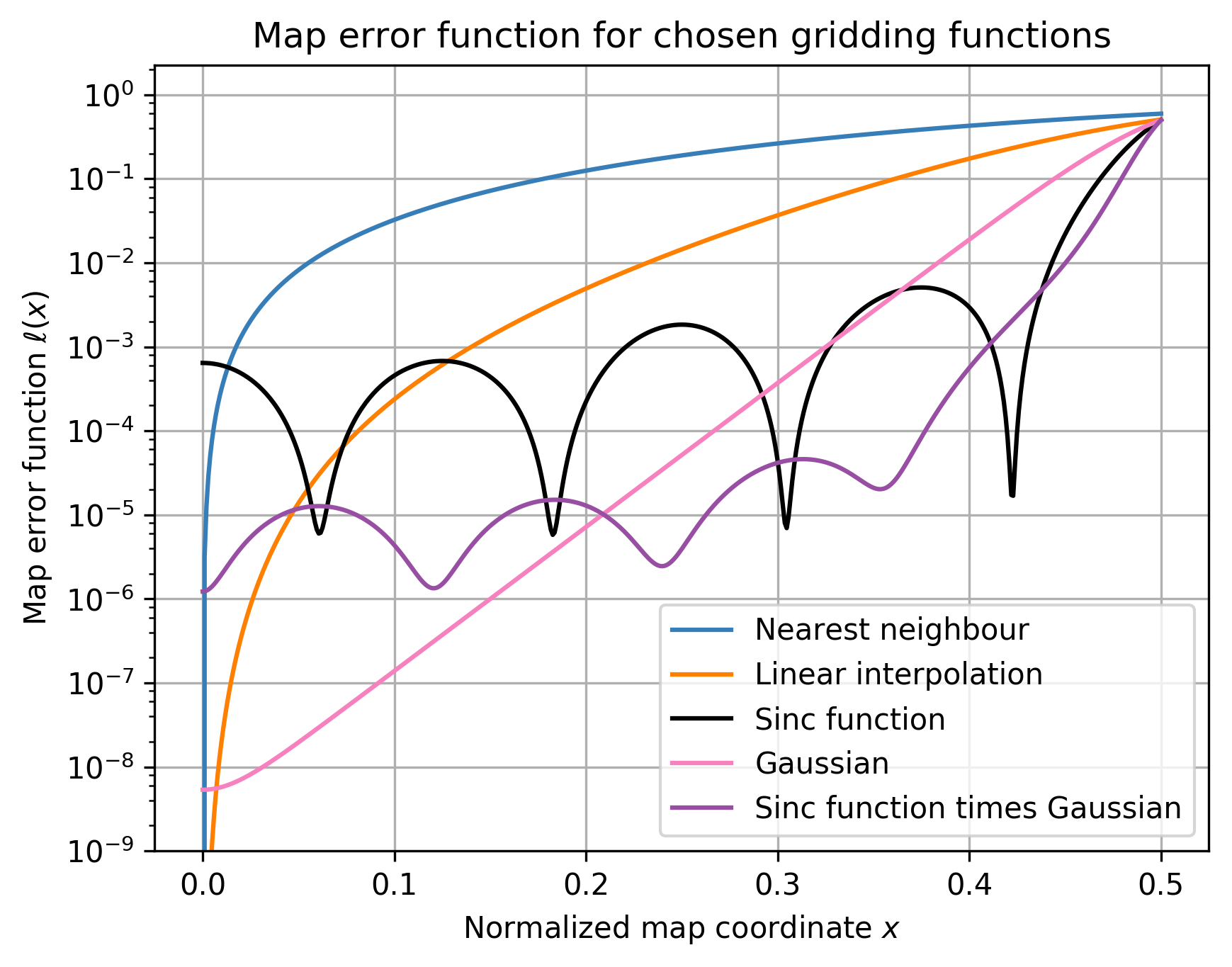}
  \caption{The map error function for nearest neighbour, linear interpolation and three other gridding functions with $W=8$.\label{fig:many_maperr_demo}}
\end{figure}

Equation (\ref{eq:l(x)}) can be used to determine the map error $\ell(x)$ for any specific choice of gridding function $C(u)$ and correcting function $h(x)$. Before presenting the results for the least--misfit gridding function, we present $\ell(x)$ for a number of common gridding functions, for the purpose of comparison. In Figure (\ref{fig:many_maperr_demo}) we show the map error function for the following five choices of gridding function $C(u)$:

\begin{enumerate}
	\item Nearest neighbour interpolation, i.e. $C_1(u) = 1$ on the interval $-0.5\leq u < 0.5$, and zero elsewhere.
	\item Linear interpolation, i.e. $C_2(u) = 1 - |u|$ on the interval $-1 \leq u < 1$, and zero elsewhere.
	\item Truncated $\sinc$ function for $W=8$, i.e., $C_3(u)=\sinc(u)$ for $-4\leq u < 4$, and zero elsewhere.
	\item Gaussian function for $W=8$, i.e., $C_4(u)\propto \exp(-|u|^2)$ for $-4\leq u < 4$, and zero elsewhere.
	\item Truncated $\sinc$ times Gaussian function for $W=8$, i.e., $C_5(u)\propto \exp{-\left(\frac{|u|}{2.52}\right)^2} \sinc \left(\frac{u}{1.55}\right)$ for $-4\leq u < 4$, and zero elsewhere.
\end{enumerate}

The constants are chosen according to the recommendations of \citet{1999ASPC..180..127B}. In each case the error becomes large at the edge of the map, $x=0.5$. The performance near the centre of the map is much better, with functions that include the $\sinc$ factor spreading out the range of $x$, over which the error is relatively small, over a larger portion of the map.

\begin{figure}
\centering\includegraphics[width=\columnwidth]{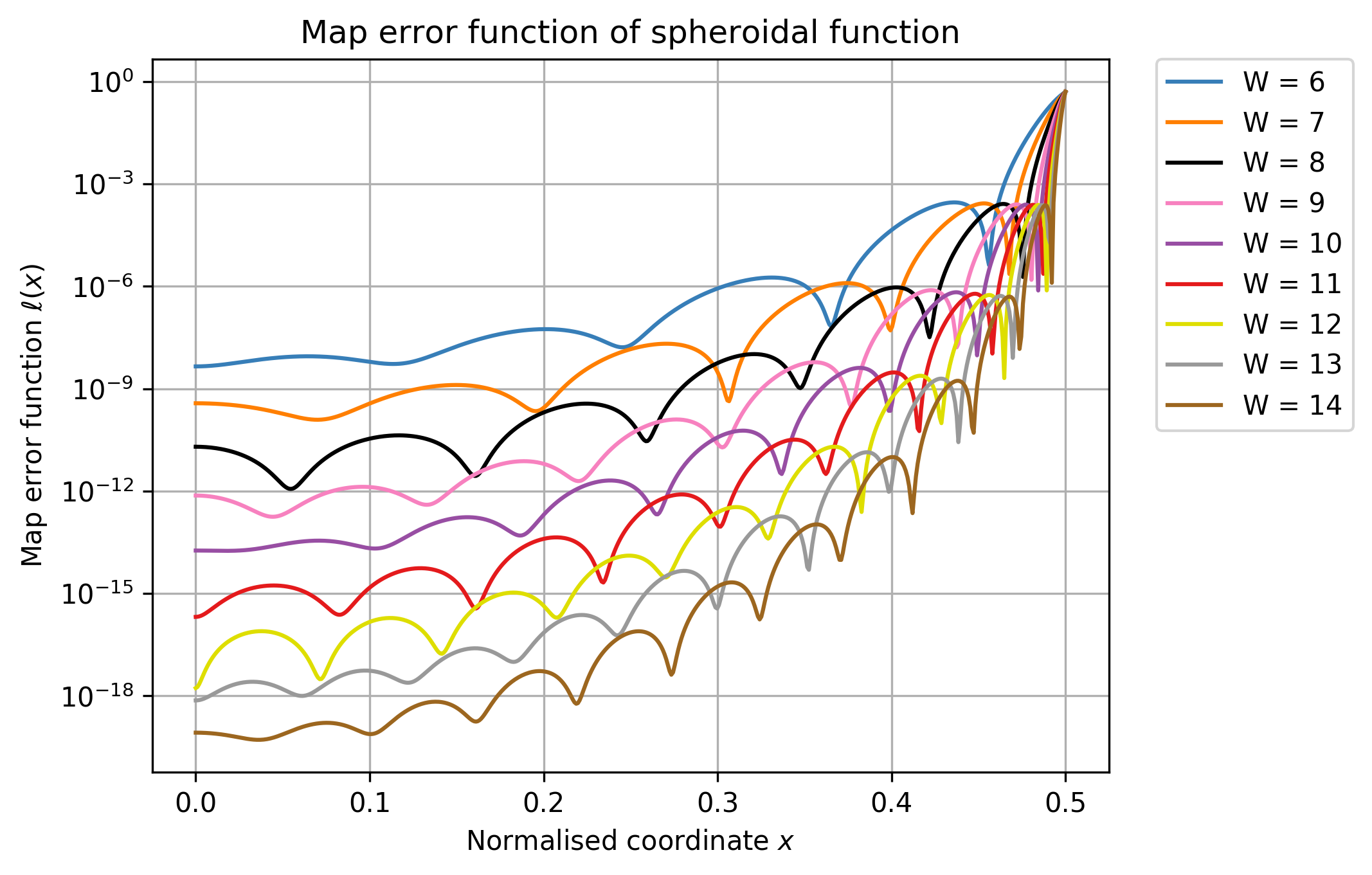}
\caption{The map error functions for spheroidal functions, with $W$ ranging from $6$ to $14$.\label{fig:Spheroidal_Function_maperr_demo_a1}}
\end{figure}

Figure (\ref{fig:Spheroidal_Function_maperr_demo_a1}) shows the map error function for the zero--order spheroidal function and values of $W$ in the range $6$ through $14$. The map errors are much smaller than for the five functions considered previously for a comparable value of $W$. In particular, comparison of the results for $C_3$, $C_4$ and $C_5$ for $W=8$ in Figure (\ref{fig:many_maperr_demo}) with the black line in Figure (\ref{fig:Spheroidal_Function_maperr_demo_a1}) reveals the clear superiority of the spheroidal function. By increasing the value of $W$ used with the spheroidal function, the error can be reduced substantially, especially over the central portion of the map. As we move away from the centre, however, the error increases until it becomes of order unity at the edge.

\begin{figure}
\centering\includegraphics[width=\columnwidth]{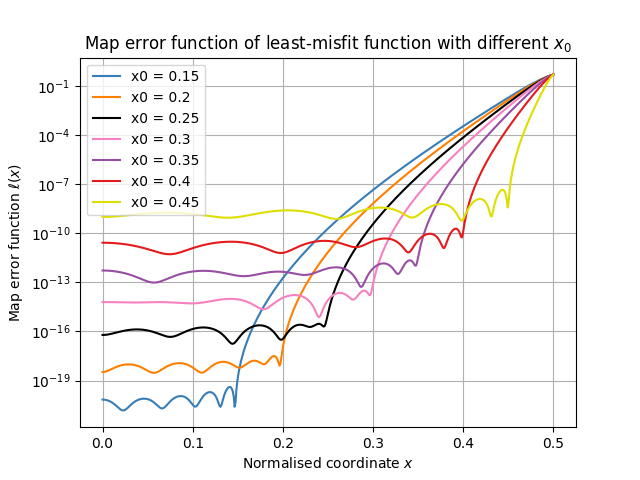}
\caption{Map error functions for the least--misfit gridding functions with different values of $x_0$, and $W=8$.\label{fig:map_error_x0}}
\end{figure}

Let us now consider the results for the least--misfit gridding functions. In Figure (\ref{fig:map_error_x0}), the map error function $\ell(x)$ optimised with different values of $x_0$ with $W=8$ generally has some features in common:
\begin{itemize}
	\item $\ell(x)$ remains fairly stable, with small fluctuations from $x = 0$ to $x=x_0$.
	\item As $x$ exceeds $x_0$, the map error function increases rapidly.
\end{itemize}

When the value of $x_0$ is reduced so as to increase the discarding range of the dirty image in Figure (\ref{fig:map_error_x0}), the error within the retained range is reduced. But the cropping of a larger portion of the image gives rise to larger FFT computations in order to make a larger dirty image. The choice of $x_0$ must therefore be considered carefully.

\begin{figure}
\centering\includegraphics[width=\columnwidth]{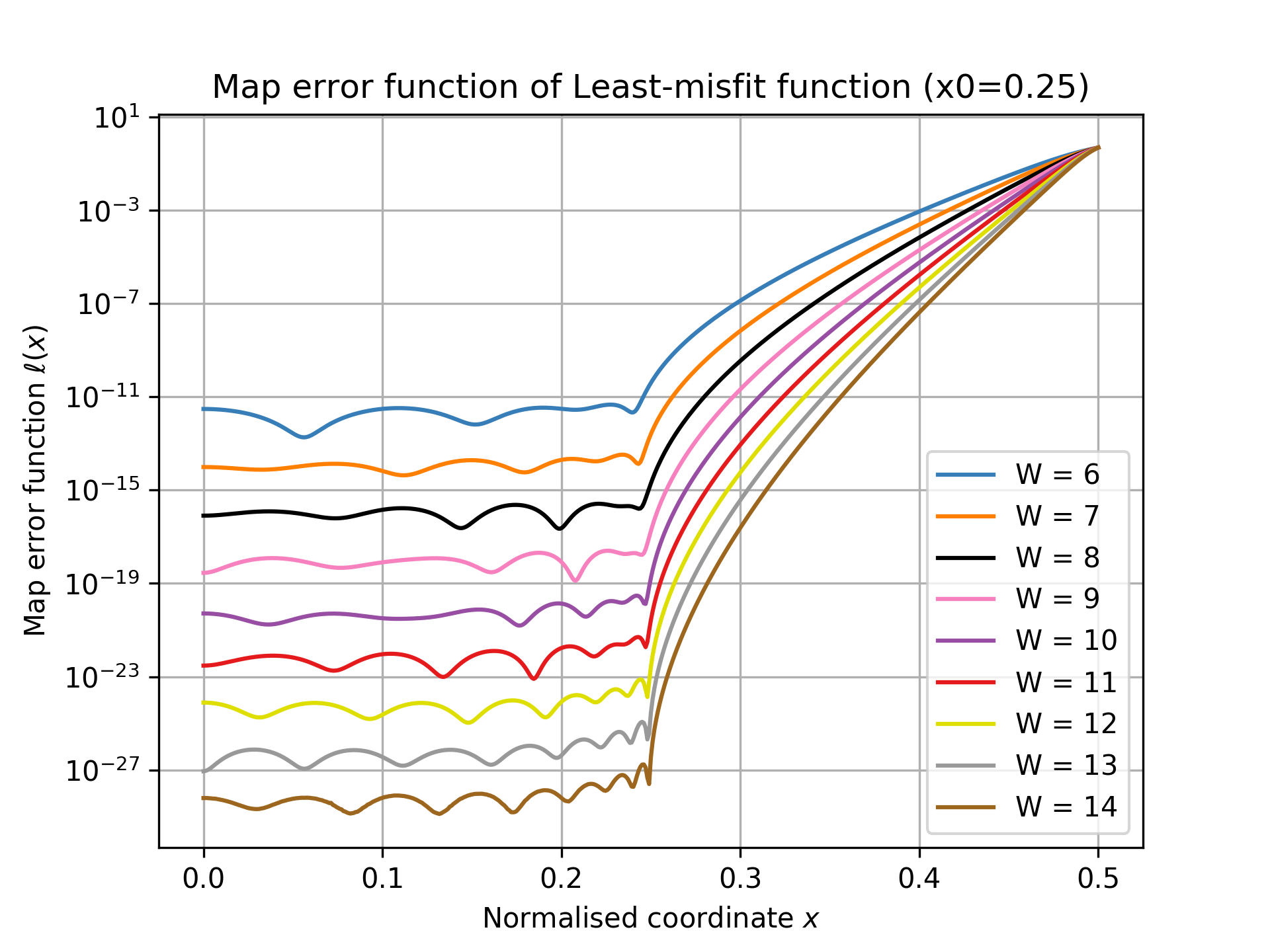}
\caption{Map error functions for the least--misfit gridding functions when $x_0 = 0.25$, with $W$ ranging from 6 to 14 using the matrix $\mathbf{B}$.\label{fig:W14demo}}
\end{figure}

We now consider the results for $x_0=0.25$, for which we seek to minimise the upper bound of the error over the central half of the map (in one dimension). Figure (\ref{fig:W14demo}) shows the map error function in this case for $W=6$ through $W=14$. To obtain the results for $W\geq 12$ in Figure (\ref{fig:W14demo}) using double precision arithmetic, it was necessary to reformulate part of optimisation algorithm in a slightly different way in order to ensure numerical stability. Details are given in Appendix \ref{app:2}. The map error function values attained in this central region are at least $100$ times better than those for the spheroidal function with the same $W$. Taking the case with $W=7$, the least--misfit function can achieve a $\ell(x)$ at $10^{-14}$, whereas the spheroidal function requires $W\geq 10$ and $x_0\leq 0.2$ for their error functions to be comparable. The improvement becomes greater as $W$ is increased.

For gridding using single-precision arithmetic, we recommend using $W=7$, since the map error (which bounds $|e(x)|^2$ in Equation (\ref{eq:e(x)2}) is less than $10^{-14}$ over $-x_0 \leq x \leq x_0$.

\section{Least--misfit gridding and correcting functions for $x_0=0.25$}

\begin{figure*}
\begin{center}
\includegraphics[width=0.4\textwidth]{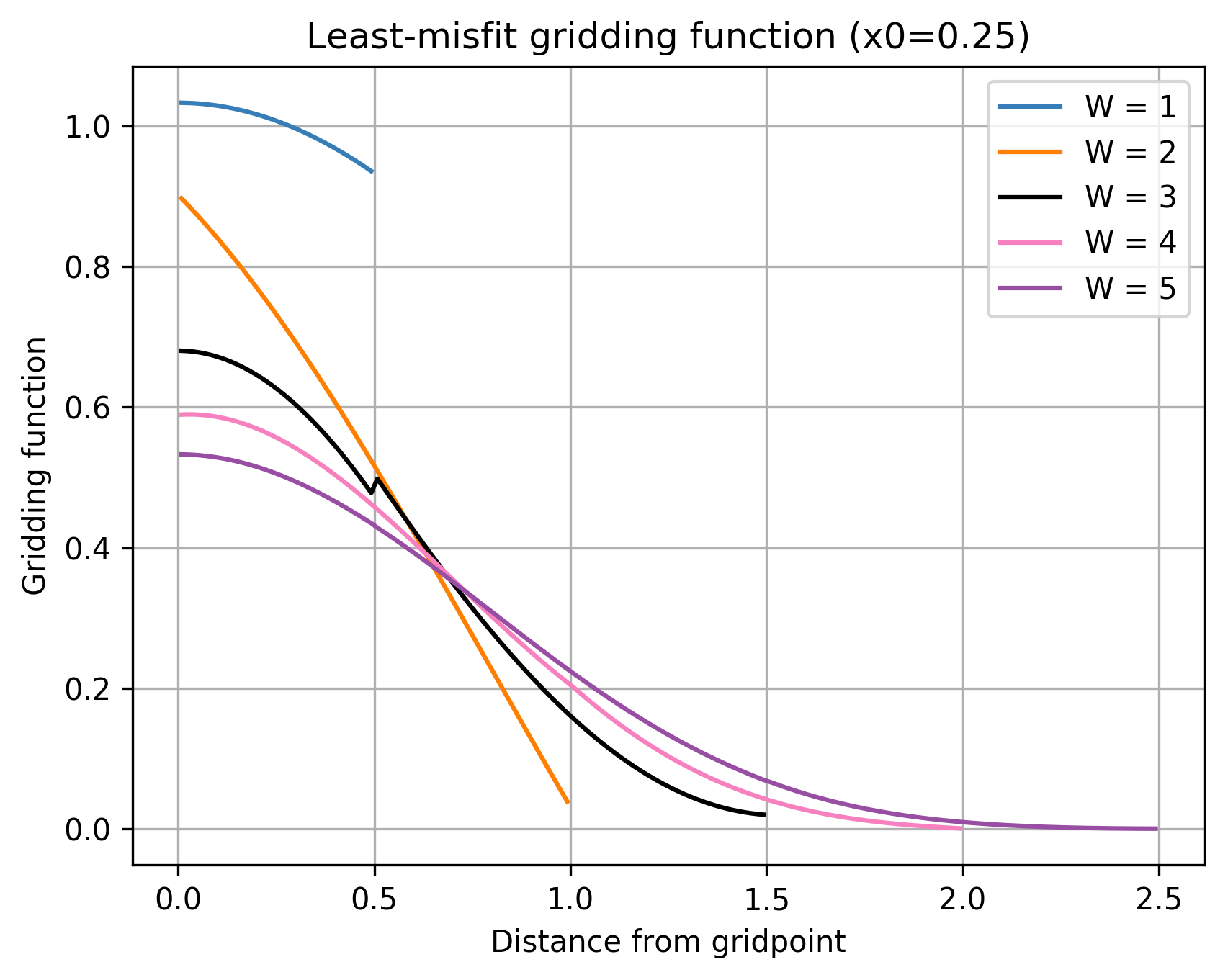}
\includegraphics[width=0.4\textwidth]{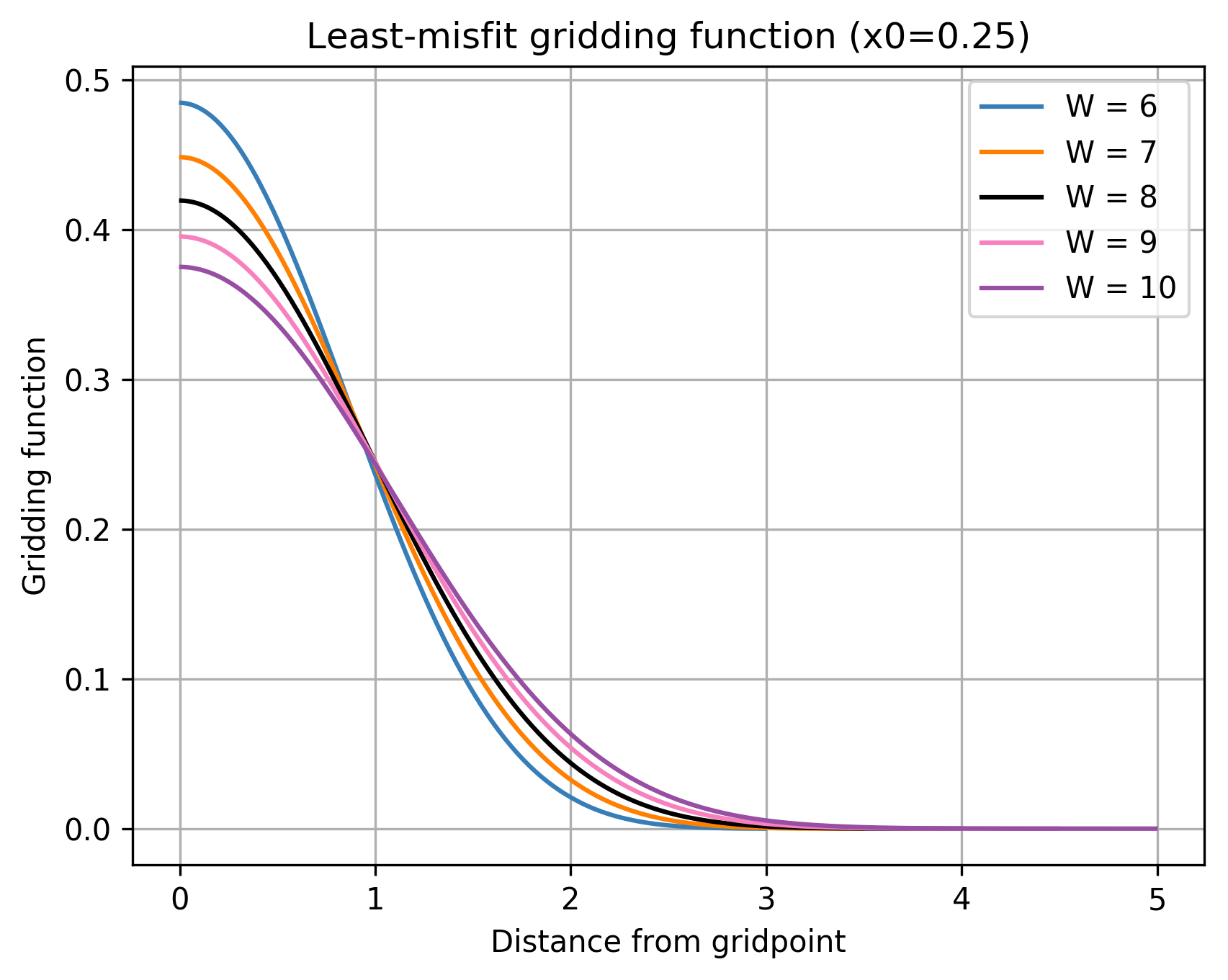}
\includegraphics[width=0.4\textwidth]{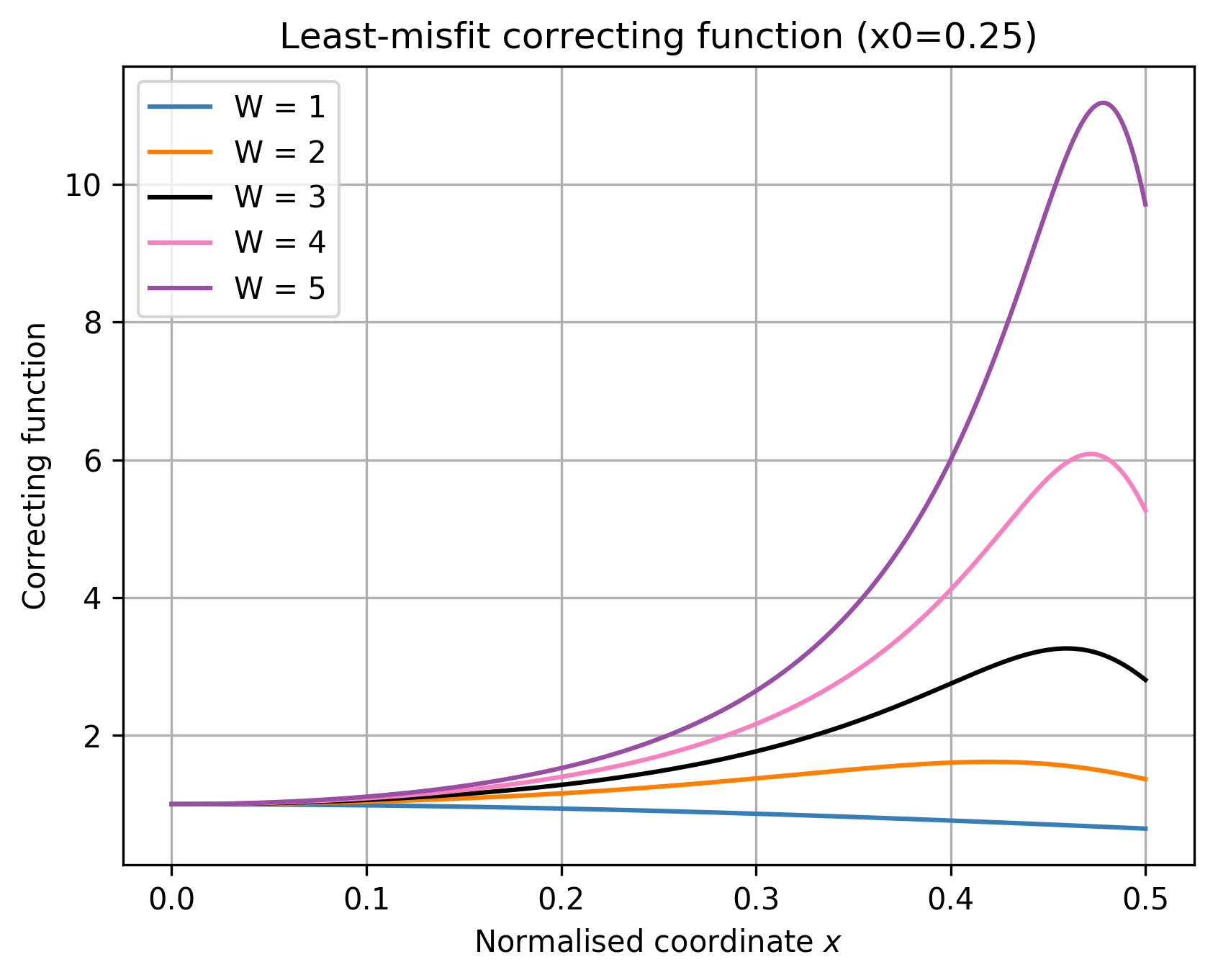}
\includegraphics[width=0.4\textwidth]{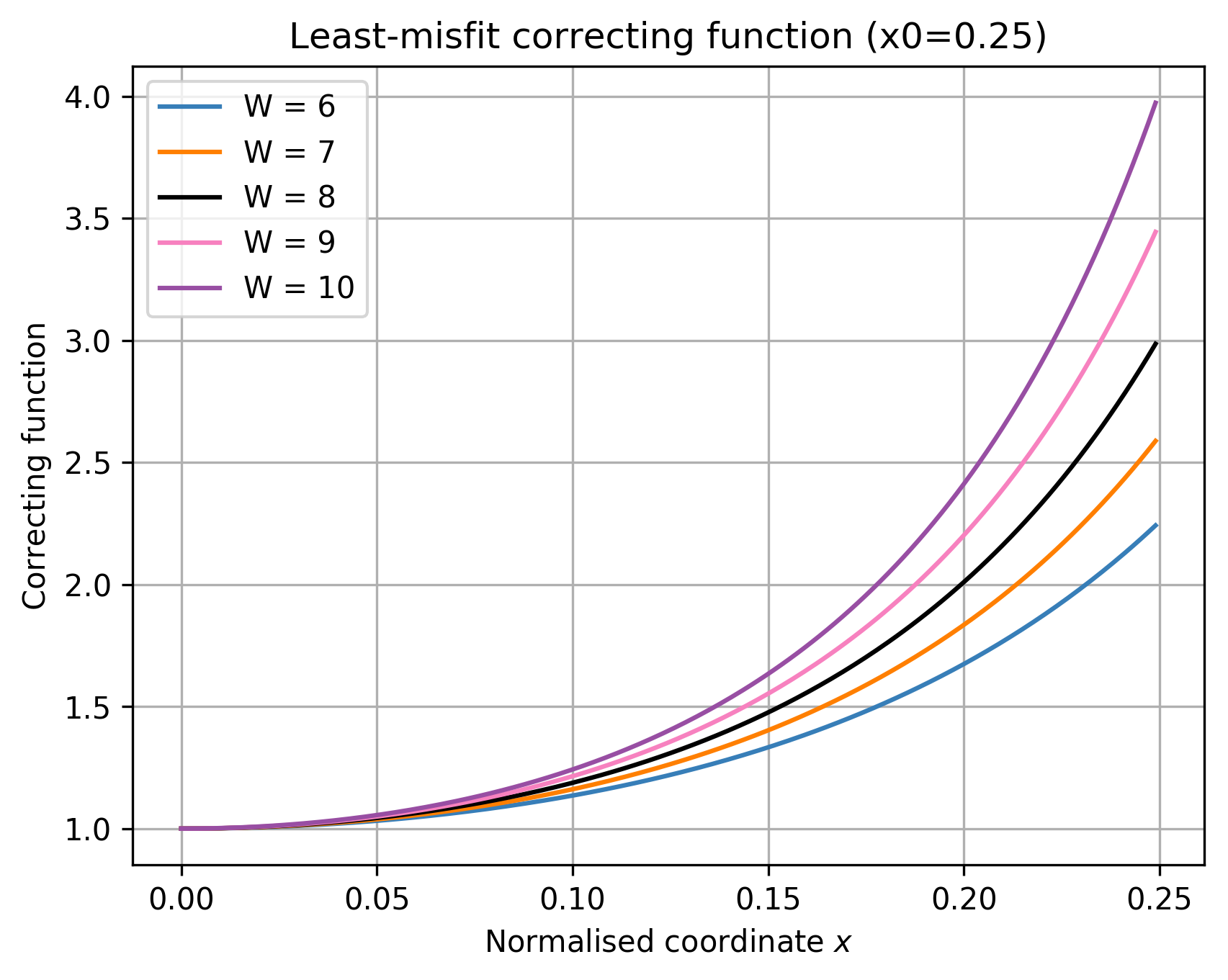}
\caption{The least--misfit gridding functions, and the corresponding correcting functions when $x_0 = 0.25$ with $W$ varying from 1 to 10.\label{fig:W15demo}}
\end{center}
\end{figure*}

In this section we show graphs of the least-misfit gridding and correcting functions for $x_0=0.25$. These are shown in Figures (\ref{fig:W15demo}) for values of $W$ ranging from $1$ to $10$.

Several interesting points are most apparent for small values of $W\leq 3$, although larger values of $W$ would typically be used in order to take advantage of the improved map error. These points are discussed in Appendix \ref{app:3}

For $W\geq 2$ the correcting function $h(x)$ increases monotonically over the range $0\leq x \leq x_0$. However, the bottom left panel of Figure (\ref{fig:W15demo}) shows that, as $x$ extends past $x_0$ and approaches the edge, $h(x)$ begins to decrease once more. This also happens for larger values of $W$, although it is not apparent from the bottom right panel, which is plotted for values of $x$ extending only up to $x_0$. 

By substituting the optimal correcting function $h(x)$ given by Equation (\ref{eq:h(x)}) into expression (\ref{eq:l(x)}) for the map error, we find that
\begin{equation}
	\ell(x) = 1-h(x)\int_0^1\mathrm{d}\nu\sum_rC(r-\nu)\cos[2\pi(r-\nu)x]= 1-h(x)c(x).
\end{equation}
In the portions of the map where $\ell(x)$ is small, $h(x)\approx 1/c(x)$, which is the most commonly used correcting function formula. When we use the least--misfit gridding functions, we retain only the portion of the map with $|x|\leq x_0$. In this central portion, the difference between $1/c(x)$ and the optimal $h(x)$ given by Equation (\ref{eq:h(x)}) or (\ref{eq:correction_func}) is negligible.

\begin{figure}
\centering\includegraphics[width=\columnwidth]{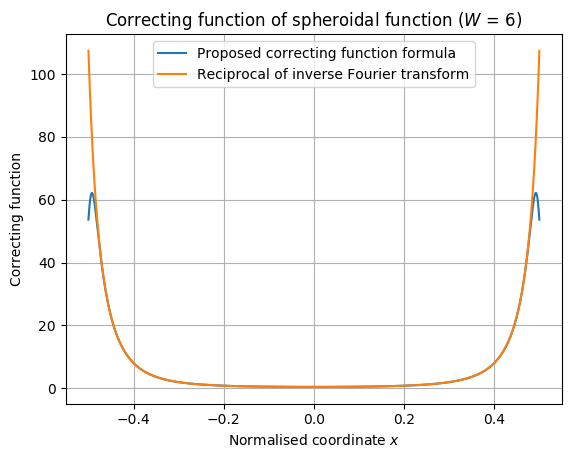}
\caption{Correcting functions for the spheroidal function ($W = 6$) calculated via Equation (\ref{eq:correction_func}) and from $1/c(x)$. The difference inside the central part of the image is as small as $10^{-8}$, almost indistinguishable. \label{fig:Schwab_different_correctingfunc}}
\end{figure}

The difference between $h(x)$ and $1/c(x)$ is significant only if $l(x)$ is not small. We plot the correcting functions for the spheroidal function with $W = 6$ using the two different formula, as shown in Figure (\ref{fig:Schwab_different_correctingfunc}). The correcting function calculated based on Equation (\ref{eq:correction_func}) is plotted as the blue line, and that calculated via $1/c(x)$ is plotted as the orange line. The difference between these becomes significant only as $x$ approaches $0.45$, where the corresponding $\ell(x)$ becomes as large as $10^{-3}$.

\begin{figure}
   \centering\includegraphics[width=\columnwidth]{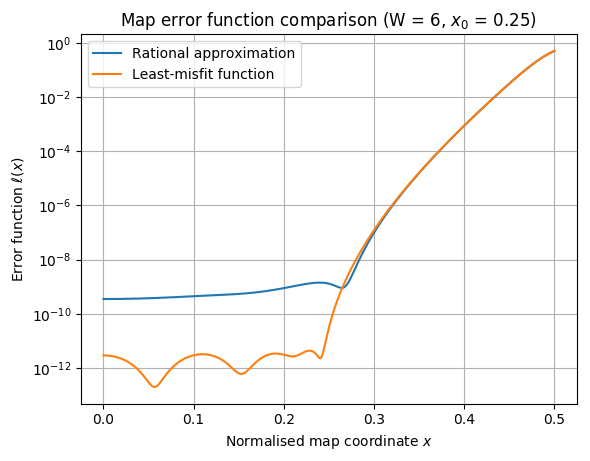}
\caption{The error function for the least--misfit gridding function ($x_0 = 0.25, W=6$) and its rational approximation respectively. \label{fig:difference_tan_rational}}
\end{figure}

We then compare the rational approximation of the gridding function proposed by \citet{tan1986aperture} using the same criteria against our least--misfit gridding function when $W = 6$ and $x_0 = 0.25$. Figure (\ref{fig:difference_tan_rational}) shows the corresponding map error functions. The least--misfit function has an error function $\ell(x)$ approximately $100$ times smaller than that of the rational approximation. Since the error function is the upper bound of the square of the image misfit, our updated implementation improves the image misfit roughly tenfold when $W=6$ and $x_0=0.25$.

\section{Comparison of least--misfit function and spheroidal function}\label{sec:c4s4}

We have already made a theoretical comparison between the least--misfit function and the spheroidal function, in our comparison of the map error function for the two gridding functions. This section reports a numerical experiment to make comparison using simulated data, studying the image misfit and the suppression of aliasing.

\subsection{Comparison of image misfit}

We simulated a VLA snapshot observing 34 $4$ GHz point sources with differing fluxes and locations across the field of view. No noise sources or other complications were added. The simulation was generated with the VLA A--array configuration. More details of this dataset are given in Appendix \ref{app:4}.

Figure (\ref{fig:Image_misfit_leastmisfit_fullmap_sqaure_testcard}) shows the RMS value of the image misfit. The gridding functions used are least--misfit functions with differing values of $W$. The $x$--axis represents the normalised image plane coordinate. For a given value of $x$, the $y$--axis is the RMS value of the image misfit within the range $[-x,x]$. This choice accurately reproduces the features found in Figure (\ref{fig:W14demo}). The RMS value of the image misfit remains at a small and steady value from $0$ to $0.25$. It then undergoes a rapid increase from $0.25$ to $0.5$, where the image will be cropped. The final FFT image therefore successfully maintains a small image misfit from its corresponding DFT image across the entire image. For this particular simulated data, the least--misfit function with $W = 7$ already causes the image misfit to reach the limit of single precision floating point arithmetic. The mild fluctuations visible in Figure (\ref{fig:W14demo}) do not manifest significantly because of the averaging operation involved in taking the RMS value.

\begin{figure}
  \centering\includegraphics[width=\columnwidth]{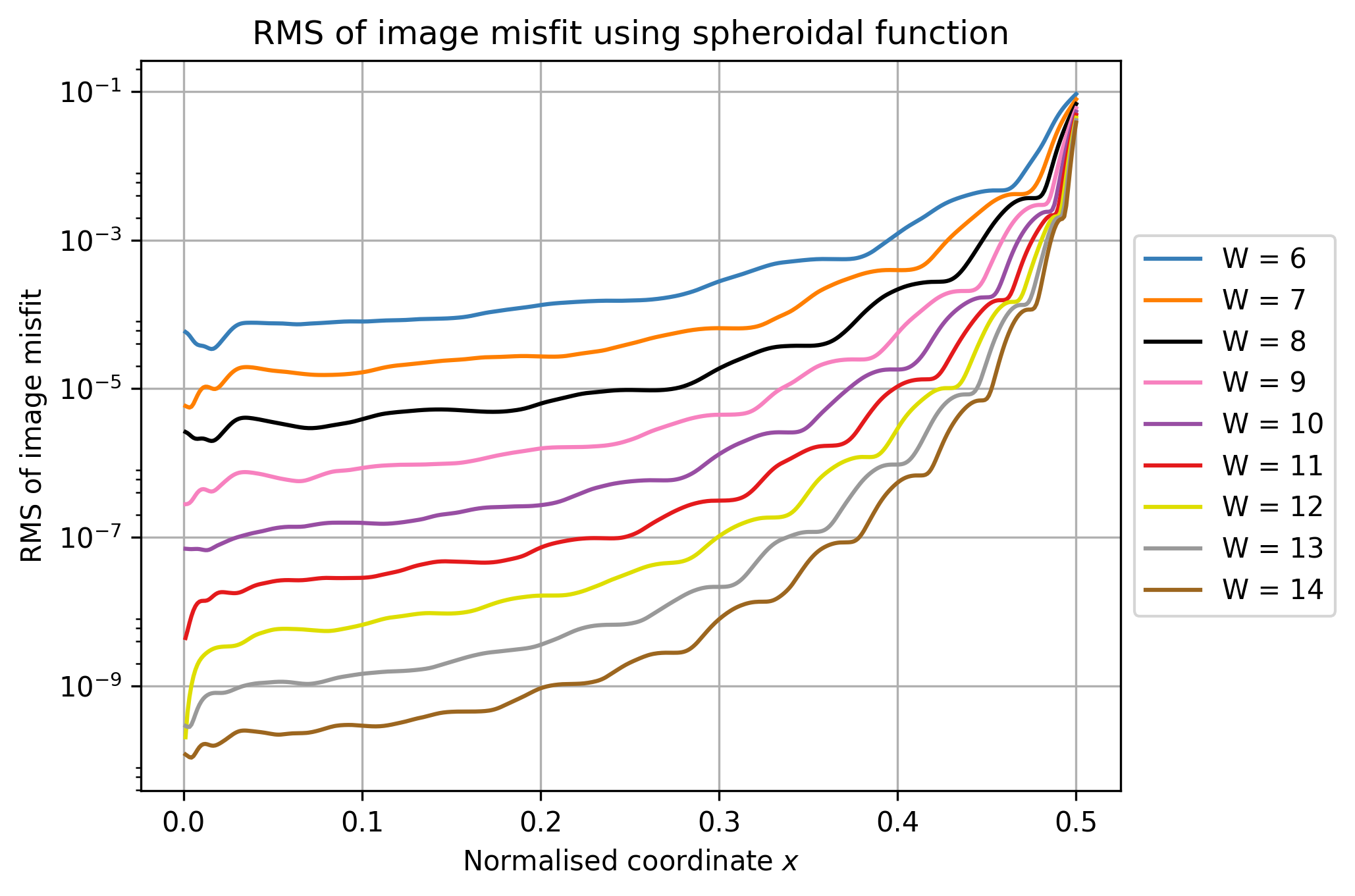}
  \caption{RMS value of the image misfit for 34 point--source simulated data using spheroidal functions ($\alpha = 1$) with differing values of $W$. \label{fig:Image_misfit_PSWF_fullmap_sqaure_a01_testcard}}
\end{figure}

Figure (\ref{fig:Image_misfit_PSWF_fullmap_sqaure_a01_testcard}) shows the RMS value of the image misfit for the same data, using spheroidal functions with different values of $W$. This figure also shares the same features as Figure (\ref{fig:Spheroidal_Function_maperr_demo_a1}).

Comparison of Figure (\ref{fig:Image_misfit_leastmisfit_fullmap_sqaure_testcard}) with Figure (\ref{fig:Image_misfit_PSWF_fullmap_sqaure_a01_testcard}) reveals that, in the range $[0,0.25]$, the image misfit using the least--misfit gridding function is at least $10^2$ times better than that using the spheroidal function with the same window width $W$. To achieve single precision in the image misfit, the least--misfit gridding function needs only a support width of $W=7$ with $x_0 = 0.25$, whereas PSWF requires a width $W=10$ with image cropping from at least $x_0=0.2$.

\begin{figure}
  \centering\includegraphics[width=\columnwidth]{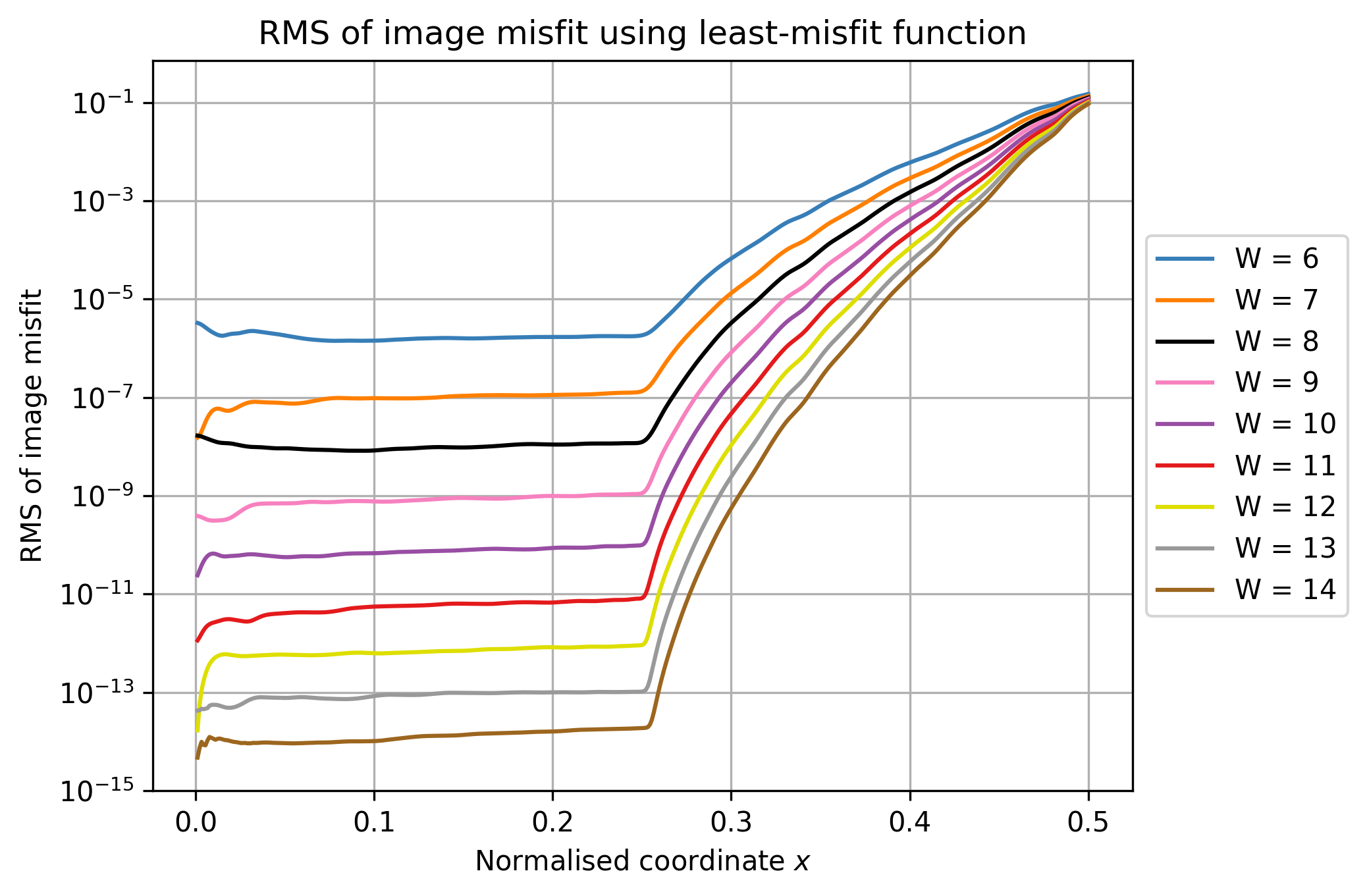}
  \caption{RMS value of the image misfit for 34 point--source simulated data using the least--misfit functions with differing values of $W$.  \label{fig:Image_misfit_leastmisfit_fullmap_sqaure_testcard}}
\end{figure}

\begin{figure}
  \centering\includegraphics[width=\columnwidth]{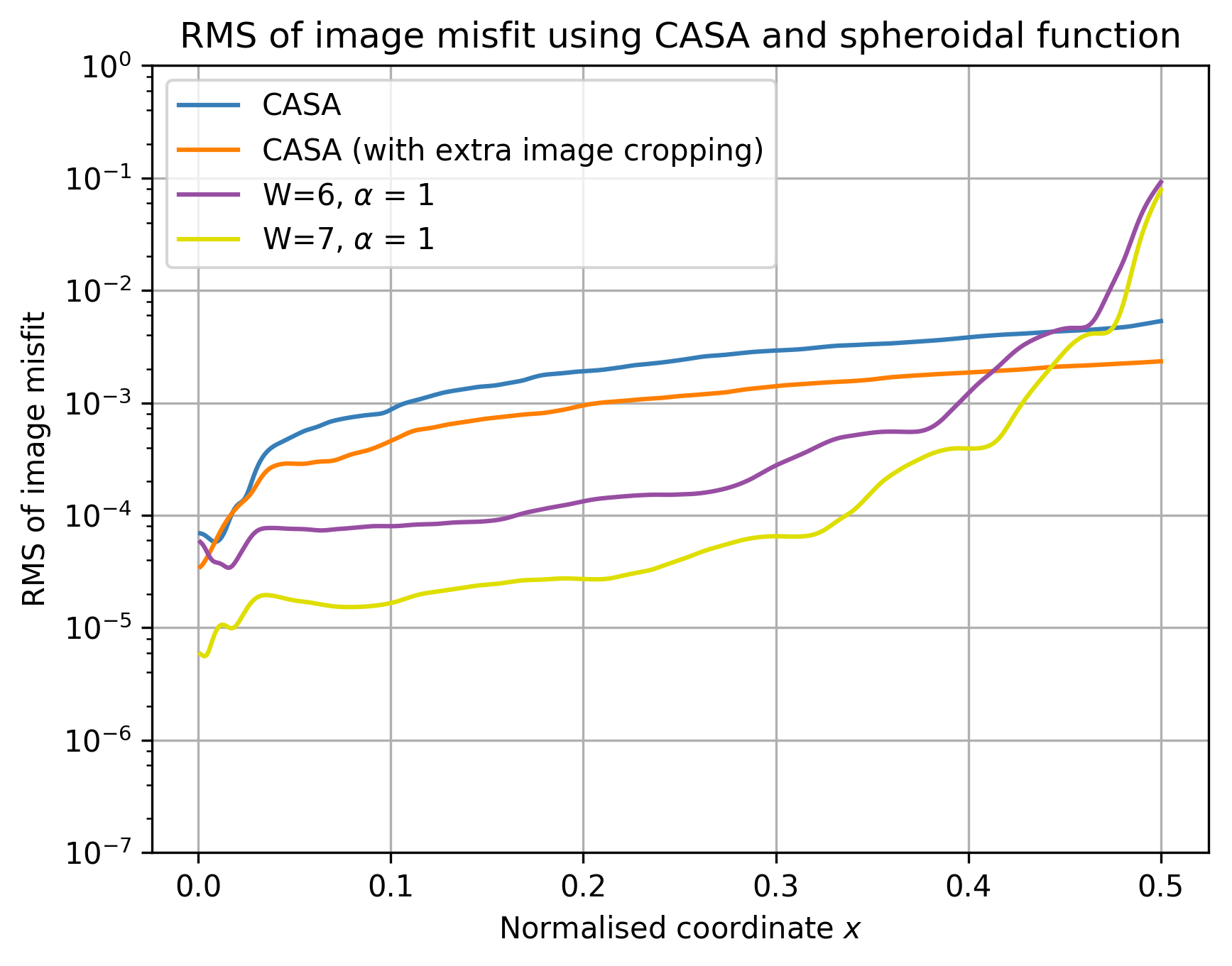}
  \caption{RMS value of the \texttt{CASA} dirty image misfit for the 34-source simulated data.  \label{fig:Image_misfit_CASA_fullmap_sqaure_testcard}}
\end{figure}

In addition, we construct a dirty image using the same data as in \texttt{CASA} for comparison with the DFT dirty image; the RMS value of the image misfit is plotted in blue in Figure (\ref{fig:Image_misfit_CASA_fullmap_sqaure_testcard}). We then make a double--sized dirty image via \texttt{CASA} with the same pixel size and crop the outer half of the image, to determine whether the image cropping influences the image misfit. The corresponding RMS values are plotted in orange. The two further lines are identical to the lines with $W=6,7$ in Figure (\ref{fig:Spheroidal_Function_maperr_demo_a1}).

Since the dirty image has already been cropped in \texttt{CASA}, the blue line in the figure attains a misfit of $10^{-3}$ at around $x=0.5$, compared to $10^{-1}$ achieved by spheroidal functions with $W=6,7$ with no cropping. Further image cropping, which corresponds to the orange line, reduces the image misfit by much less than $10$. It is therefore unnecessary to perform extra image cropping on \texttt{CASA} dirty images. To achieve the same image misfit, we need only use $W=3$ for the least--misfit gridding function, leading to a much lower gridding computational cost.

\subsection{Comparison of aliasing suppression performance}

We consider a simple aliasing scenario in which there is no source within the field of view, and a single point source outside it which causes aliasing within it. We simulate a set of point--source visibility data based on the VLA A--array configuration, and shift the point source from very close to the edge to further away. The point source was kept outside the field of view, and its horizontal distance from the image edge was varied but with its declination kept constant. The brightness of the alias is expected to decrease as a result of the aliasing suppression using the gridding function.

\begin{figure}
  \centering\includegraphics[width=\columnwidth]{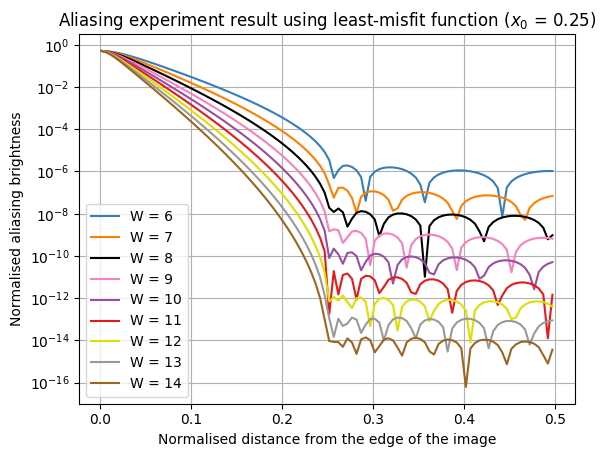}
  \caption{Normalised brightness of alias versus the horizontal distance from the point source to the edge of the field of view. Least--misfit functions with differing values of $W$ are used. The $x$--axis represents the normalised distance from the point source to the image edge, and the $y$--axis is the normalised aliasing brightness.
  \label{fig:Aliasing_comparison_leastmisfit_r2345}}
\end{figure}

We constructed DFT and FFT dirty images from the datasets, and recorded the brightness of the aliases from the images of the differences. Figure (\ref{fig:Aliasing_comparison_leastmisfit_r2345}) shows the normalised brightness of aliases versus the horizontal distance in the normalised coordinate $x$ from the source to the edge of the image. The least--misfit functions are used with different values of $W$.

The trend in the aliasing suppression shown in Figure (\ref{fig:Aliasing_comparison_leastmisfit_r2345}) is consistent with the map error functions for the least--misfit gridding functions in Figure (\ref{fig:W14demo}). Aliases in the central half of the image are well suppressed, with a normalised brightness of the alias of approximately $10^{-7}$ for $W=7$. The small fluctuations of the normalised brightness within $[0,0.25]$ are also consistent with Figure (\ref{fig:W14demo}).

\begin{figure}
  \centering\includegraphics[width=\columnwidth]{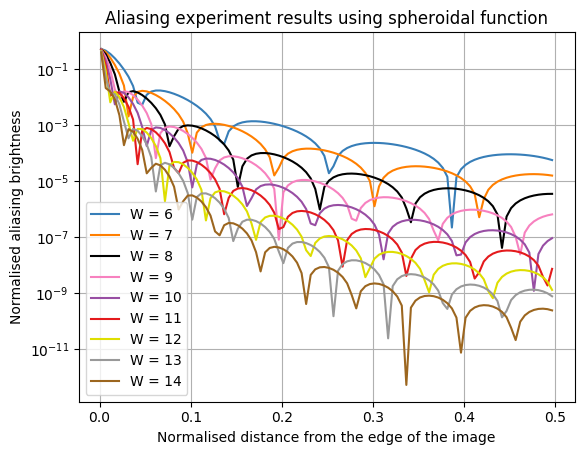}
  \caption{Normalised brightness of the alias versus the horizontal distance from the point source to the edge of the field of view. The gridding functions used here are the spheroidal functions with differing values of $W$. The $x$--axis represents the normalised distance from the point source to the image edge, and the $y$--axis is the normalised aliasing brightness.\label{fig:Aliasing_comparison_PSWF_a1_r2345678}}
\end{figure}

As expected, the aliasing effect within the outer half of the image deteriorates as the distance between the source and the field edge decreases. Furthermore, as $W$ increases, the least--misfit function suppresses aliasing more strongly. For $W\geq 7$, the brightness of the alias is already less than $10^{-7}$ of the original brightness within the range $[0,0.25]$.

The same numerical experiment was then repeated using spheroidal functions ($\alpha = 1$) with different values of $W$. Figure (\ref{fig:Aliasing_comparison_PSWF_a1_r2345678}) shows the results. This is consistent with the results in Figure (\ref{fig:Spheroidal_Function_maperr_demo_a1}): the image misfit in this case is the normalised aliasing brightness, and it grows with small fluctuations as the distance from the image centre increases. As $W$ increases, the normalised aliasing brightness reduces. For $W\geq 10$, the alias brightness is at least $10^{-7}$ of the original brightness within the range $[0,0.25]$.

With $W$ fixed, the least--misfit gridding function achieves at least $10^2$ times smaller normalised aliasing brightness than the spheroidal function at the same position. We conclude that the least--misfit gridding function is clearly superior at suppressing aliasing than the spheroidal function with the same value of $W$.

Given its good performance in aliasing suppression, the least--misfit gridding function can potentially replace the spheroidal function in wide--field imaging algorithms such as the $w$--projection method \citep{Cornwell2003W} and $w$--Stacking method \citep{humphreys2011analysis, 2014MNRAS.444..606O}. We are writing a paper proposing an improved $w$--Stacking method, in which our least-misfit gridding functions are used along with modifications on the original $w$--Stacking method. As a result, we can make the difference between the DFT and FFT dirty images for wide-field observations negligible to single precision by using $W = 7$ and to double precision by using $W = 14$.

\section{Degridding with the least--misfit gridding function}\label{sec:degridding}

The quality of a degridding procedure may be evaluated in terms of an RMS `visibility misfit', which is defined as the difference between the original visibilities and those degridded from an image model. In the following numerical experiment, we reuse the simulated data for 34 point sources and the VLA A-array described in Appendix \ref{app:4}.

\begin{figure}
  \centering\includegraphics[width=\columnwidth]{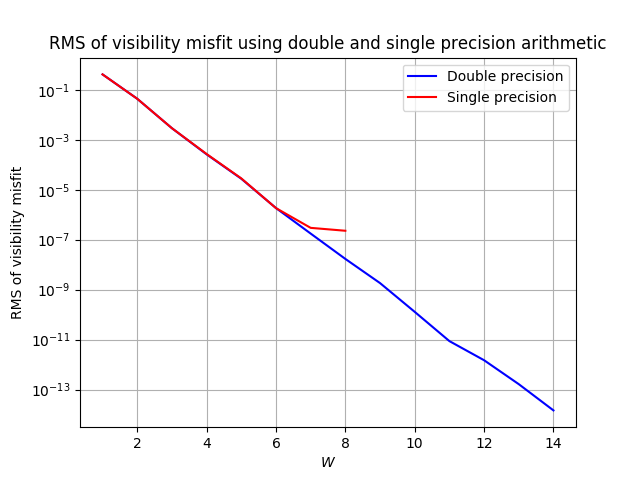}
  \caption{RMS value of the degridding misfit when using the least--misfit gridding function with different values of $W$. \label{fig:degridding_plot}}
\end{figure}

Figure (\ref{fig:degridding_plot}) shows the results for the least-misfit gridding functions with $x_0=0.25$ for different values of $W$, using both single and double precision arithmetic. For single precision arithmetic, there is no advantage in using $W>7$.

In the Cotton-Schwab CLEAN algorithm \citep{1983AJ.....88..688S}, the visibilities are calculated from the current CLEAN model and are subtracted from the measured visibility data during each major cycle in order to obtain a set of visibility residuals which are gridded to form a new dirty map for minor cycles of cleaning. Ideally the calculation of the visibilities from the model is done using the DFT, but it is more computationally efficient to use an FFT followed by degridding. We have conducted experiments which confirm that by using the least-misfit function with $W=7$, the error introduced by the degridding is negligible at the level of single-precision arithmetic. In this way, no extra error is introduced during major cycles. Under the same conditions, we have found that the spheroidal function with the same support, leads to an RMS misfit at least 100 times that of the least-misfit function.

In summary it is recommended that the least-misfit gridding function with $W=7$ and $x_0=0.25$ be also used for degridding when single precision arithmetic is adequate.

\section{Practical implementation of the least--misfit function}\label{c4sec:computaional_cost}

There are two distinct ways to use the least--misfit gridding function during imaging: either calculate the exact function values directly for the given visibility data, or retrieve values from a pre--calculated look--up table. In the latter case the look--up table usually has very limited samples, so the values retrieved are approximations to the exact values. We shall first discuss the use of the look--up table, with attention to the choice of the sampling number and the interpolation method. We then examine the method of direct calculation, and compare the computational cost between these two implementation methods. After that, we explore different combinations of $W$ and $x_0$ for the least--misfit function, focussing on the balance between the computational cost and the desired level of accuracy. As a result, we recommend use of the least--misfit gridding function with $W = 7$ and $x_0 = 0.25$ in order to achieve both image and visibility misfit at single floating point precision. To attain the same precision using spheroidal functions, much greater computational and storage costs would be incurred.

\subsection{Look--up table implementation}

In previous experiments, exact values of the gridding functions were calculated and used, but in practice a pre--calculated look--up table is commonly used, in which the gridding function is uniformly sampled by a so--called `sampling rate' within each grid. The bigger the sampling rate is, the finer the gridding function is sampled. Given the visibility data, the corresponding gridding weights can then be retrieved from the table using the nearest--neighbour rule. The process of generating the look--up table is set out in Appendix \ref{app:5}. 

\begin{figure}
  \centering\includegraphics[width=\columnwidth]{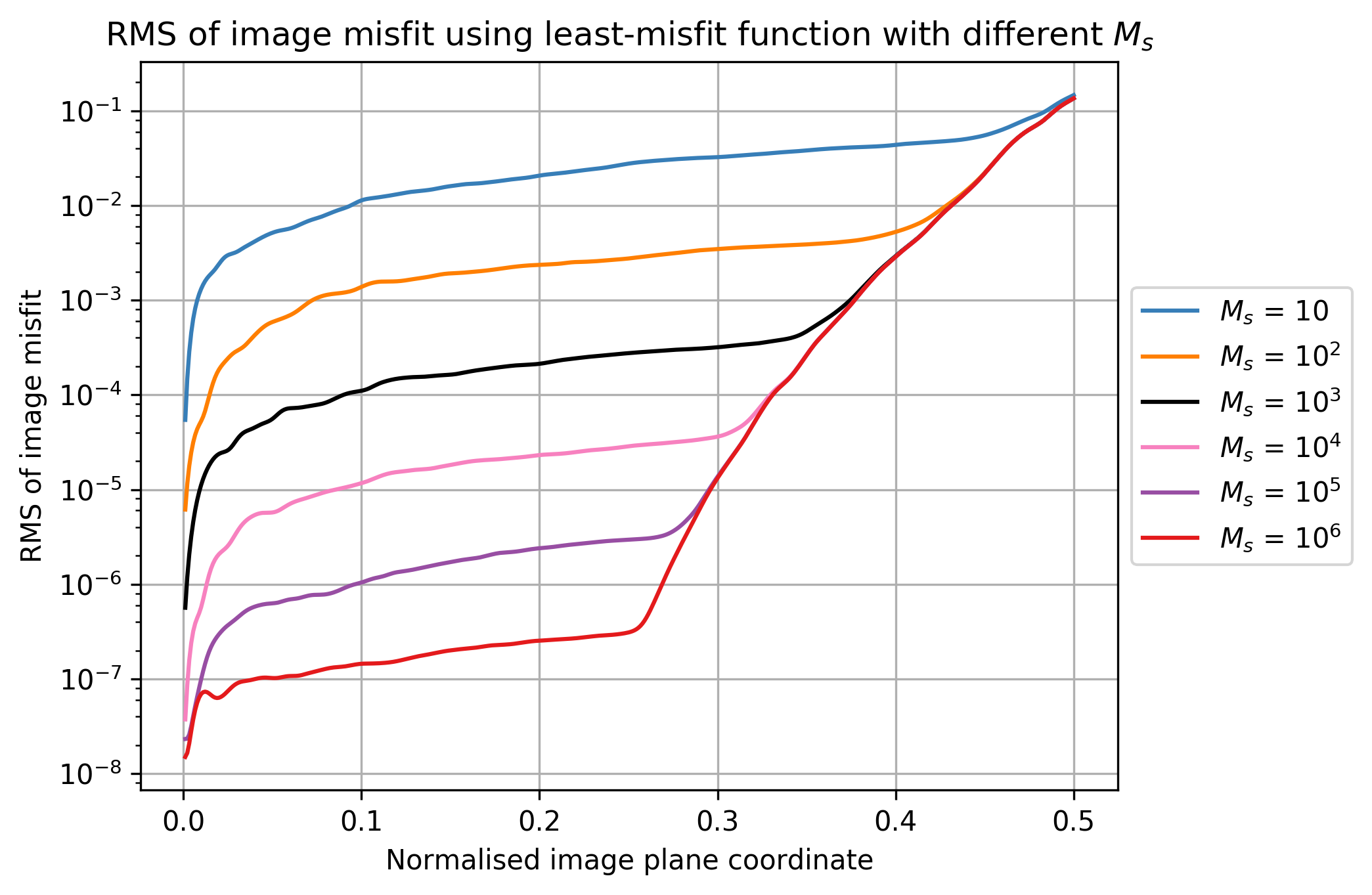}
  \caption{RMS value of the image misfit achieved using look-up tables of the least--misfit function ($W = 7$, $x_0 = 0.25$) with differing sampling rates.\label{fig:Leastmisfit_W7_nearest}}
\end{figure}

When the nearest--neighbour rule is in use, for the least--misfit function with $W=7$ and $x_0=0.25$, a sampling rate of $10^6$ already causes both the image and visibility misfit to reach the single precision limit during the gridding and degridding process; see Figure (\ref{fig:Leastmisfit_W7_nearest}) and Table (\ref{tab:degridding_W7}). The sampling rate $M_s$ changes from $10$ to $10^6$; when $M_s = 100$, the accuracy achieved is already comparable to that achieved by \texttt{CASA}, as demonstrated in Figure \ref{fig:Image_misfit_CASA_fullmap_sqaure_testcard}.

Using the same image model of the 34 point-source VLA simulated data as in Section \ref{sec:degridding}, we recorded the RMS of the difference between the degridded and original visibility in Table (\ref{tab:degridding_W7}). Although the small sampling rate prevents the gridding function from achieving its best performance, with $M_s = 10^6$ the accuracy achieved ($10^{-7}$) already matches the use of the exact gridding function values.

\begin{table}
\centering
 \caption{RMS value of the visibility misfit using the least--misfit gridding ($W=7$, $x_0=0.25$) look--up table with differing sampling rates $M_s$}
 \label{tab:degridding_W7}
 \begin{tabular}{lcc}
  \hline
   $M_s$ & RMS of the visibility misfit\\
  \hline
  $10^2$ & $6.23\times10^{-3}$\\
  $10^3$ & $4.48\times10^{-4}$\\
  $10^4$ & $4.40\times10^{-5}$\\
  $10^5$ & $4.62\times10^{-6}$\\
  $10^6$ & $4.85\times10^{-7}$\\
  No sampling & $2.98\times10^{-7}$\\
  \hline
 \end{tabular}
\end{table}

With the look--up table in use, the RMS image misfit builds up away from the centre of the map; this can be seen in Figure (\ref{fig:Leastmisfit_W7_nearest}), and also in Figure (\ref{fig:Image_misfit_CASA_fullmap_sqaure_testcard}) for which a look--up table of the spheroidal function is used in \texttt{CASA}. It is particularly obvious closer to the centre of the image. As an example, with $M_s$ fixed, the RMS of the image misfit increases rapidly from $x=0$ to approximately $x=0.02$, and then there is a more steady and gentle rise to $x=x_0$. The effect is explained in Appendix \ref{app:6}.

Compared to other interpolation methods, such as linear, quadratic or cubic interpolation, nearest--neighbour requires less memory access and lower calculation cost. It has one obvious disadvantage, however: a fairly large sampling rate $M_s$ is necessary to reach the desired accuracy. For example, when $W$ increases to $14$, a sampling rate of more than $10^{12}$ is necessary to achieve its best gridding and degridding performance. In contrast, the sampling rate can be decreased to around $10^6$ if linear interpolation is used.

\subsection{Direct calculation implementation}

The direct calculation method provides exact gridding function values. Given the visibility data, we can calculate the fractional offset part of each $u$ as $\nu = u - \lfloor u \rfloor\in [0,1]$ (in the one-dimensional case); the corresponding $W$ gridding weights can then be calculated via Equation (\ref{eq:grid_corec_eq2}). The calculation need be performed only once for a certain visibility data.

We compare the usage of the two different methods here. The look--up table only needs to be built up once and then stored. A sampling rate of $10^6$ already causes the image and visibility misfit level to attain the single precision limit using the nearest neighbour rule with $W=7, x_0=0.25$, and to attain the double precision limit using linear interpolation with $W=14, x_0=0.25$. Because of the symmetry of the gridding function it is only necessary to store $3.5\times10^6$ single precision floating points; only $13.35$MB is required to attain the single precision limit with $W=7, x_0=0.25$. On a computer with the quad--core Intel i5-2310 @ 2.90GHz processor and 8 GB of memory, such a table is constructed in $12$ seconds, corresponding to $12$ microseconds to calculate each gridding value.

For $W=14, x_0=0.25$, the corresponding table would cost $53.41$MB in storage, with double precision floating points stored. For $W>11$, the $\textbf{B}$ matrix must be calculated for each value of $\nu$, and it is therefore advisable to use the look--up table via linear interpolation to save the large direct computational cost: the look--up table with the same sampling rate and value of $W$ takes $104$ seconds to determine on the same computer, or $104$ microseconds for each gridding value.

When a look--up table is used then, for a specific visibility dataset, the nearest gridding function value must be searched for every $(u,v)$ coordinate when using the nearest--neighbour method. In our program, only $0.38$ microseconds is required on average to find $7$ indices from the look--up table for each item of visibility data. For linear interpolation, apart from the searching cost, the interpolation process generates extra computational cost, which is very much affordable. In comparison, the gridding values calculated via the direct calculation method can usually only be reused for the same visibility data, but no extra searching or interpolation cost is needed.

\subsection{Choice of $W$ and $x_0$}

No matter which method is used to obtain the least--misfit function values, the gridding process is the same. Given the number of $(u,v)$ positions specified as $N_v$, when each visibility comprising both real and imaginary parts is to be gridded onto $u$ or $v$ alone, there will firstly be 2 multiplications with the corresponding real weight. Then the weighted visibility will have $2W$ multiplications with the gridding values, and $2W$ additions on the corresponding grid points. For a two--dimensional gridding on $u$ and $v$ for given visibility data, there are $(4W^2+2) N_v$ operations. If the one-dimensional gridding function is precomputed, an extra $W^2$ multiplications will be added to find the coefficients, making the operations increase to $(5W^2+2) N_v$.

We consider next the FFT cost. Given the image pixel size $N_x$ and $N_y$, since the FFT image will be cropped, it is necessary to make a $N_x/(2x_0)$ by $N_y/(2y_0)$ FFT dirty image. Here, $y_0$ is usually equal to $x_0$. We simulated a $6$--hour long observation using the same 34-source configuration, and tested it on our computer with a quad--core Intel i5-2310 @ 2.90GHz processor and 8 GB of memory. This computer took $1.112$ seconds to compute the FFT dirty image of size 2048 by 2048, effectively taking $62$ nanoseconds to compute the FFT value on each pixel with $x_0 = 0.25$. For comparison, the gridding operation for this data only took $0.566$ seconds. The correcting function can be calculated very easily once the image size is given.

\begin{figure}
  \centering\includegraphics[width=\columnwidth]{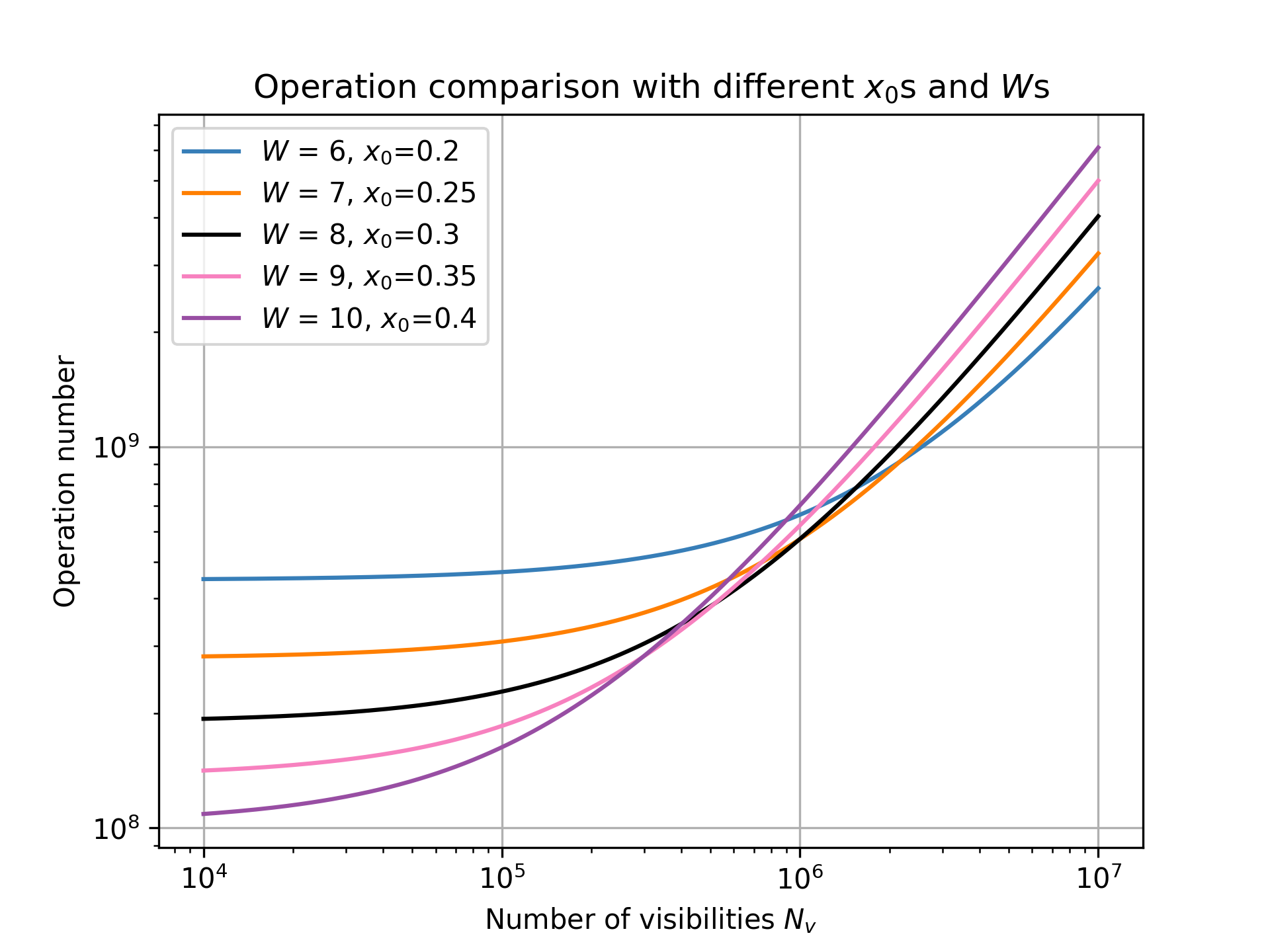}
  \centering\caption{Computational cost comparison with different combinations of $W$ and $x_0$ when using the least--misfit function. \label{fig:computational_cost}}
\end{figure}
To achieve greater accuracy in both the gridding and degridding process, $W$ should be larger and $x_0$ smaller. A larger $W$ incurs greater computational cost of the convolution, however, and a smaller $x_0$ incurs higher FFT computational costs. To achieve the desired precision at an affordable computational cost, a compromise is necessary.

The combination $W=7$ and $x_0 = 0.25$ is not the only one to attain the desired single precision. We choose 5 sets of $W$ and $x_0$ pairs for which the image and visibility misfit reach the single precision limit. The convolution and FFT operations are combined as $N_v (5W^2+2)+5\frac{N_xN_y}{4x_0y_0}\log\left(\frac{N_xN_y}{4x_0y_0}\right )$. It is assumed that $x_0 = y_0$.

The computing cost is plotted as the number of operations in Figure (\ref{fig:computational_cost}), with the visibility number ranging from $10^4$ to $10^8$ and $N_x=N_y=2048$. For $N_v>10^7$, $W$ largely dominates the overall computational cost. For the choice $W=6, x_0=0.2$, which is plotted in blue, then although the corresponding FFT image has to be cropped more than any other $W$ and $x_0$ pair, its computational cost is the smallest, because $W$ is least. When $N_v$ is smaller, $x_0$ dominates the computational cost. The computational cost of using $W=8,x_0=0.3$ and $W=7,x_0=0.25$ remain relatively small for the given range of visibility numbers. We prefer to use $W=7,x_0=0.25$, to give a smaller look--up table and setup table. Based on a good balance between the accuracy and the computational cost, we recommend using $W = 7$ and $x_0 = 0.25$ so as to achieve the single floating point precision limit in both the image and visibility misfit.

When we compare the least--misfit function against the spheroidal function, we are forced, in order to achieve the same level of image or visibility misfit, to choose a much larger $W$ for the spheroidal function; this incurs extra computational and storage costs. If we choose the same value of $W$ for both gridding functions, better precision is always obtained by using the least--misfit function. The least--misfit function is therefore preferred.

Since the computational cost saved by using a smaller value of $W$ increases with the number of visibilities, especially in view of the construction of the next generation of radio interferometers such as SKA (Square Kilometre Array)\citep{hall2008square}, it is worth using the least--misfit gridding functions with less computational cost but much higher accuracy in both the gridding and degridding processes.

\section{Conclusions}\label{sec:c4s6}

In this paper we have proposed a new set of gridding functions based on the criterion that the difference between the DFT and FFT dirty images should be minimised. As a result, accurate dirty images promise to provide more accurate information about the radio sky, and high--precision dirty images will benefit future applications that work directly with dirty images such as \citet{hague2018bayesian}.

In summary, the least--misfit gridding function outperforms the widely-used spheroidal function $100$-fold in terms of the image misfit with the same $W$. In the suppression of aliasing, even though the spheroidal function was chosen specifically to suppress aliasing effects, the least--misfit function is able to suppress the alias at least $100$ times more effectively. Essentially, aliasing is a part of the image misfit, and there is therefore no need to consider its suppression separately.

We have also considered the computational cost and the degridding process. We recommend the use of the least--misfit gridding function with $W = 7$ and $x_0 = 0.25$, so as to achieve a balance between the desired accuracy and the computational cost in both the gridding and degridding processes. If a look--up table is used, a sampling rate of $10^6$ already causes the image and visibility misfit level to reach the single precision limit using the nearest neighbour rule with $W=7, x_0=0.25$, and to reach the double precision limit using the linear interpolation with $W=14, x_0=0.25$. Such a table with $W=7, x_0=0.25$ requires only $14$M storage. If even single precision is not required, the least--misfit function can achieve the same image accuracy as \texttt{CASA} with $W=3$ and a sampling rate of $100$.

The least--misfit gridding function has a strong potential to be used to make more accurate wide--field images,

\section*{Acknowledgements}

We would like to thank John Skilling for helpful comments and Anton Garrett for his careful proofreading. This research was prompted by discussions in 1985 at the VLA between one of us (Stephen Gull) and Eric Greisen and Fred Schwab.







\appendix

\section{Algorithm for numerical optimisation}\label{app:1}

In \citet{tan1986aperture}, the minimisation of $E$ was undertaken by a joint optimisation over both $C(u)$ and $h(x)$. In the present paper we use a different approach, in which we explicitly vary only $h(x)$ in the optimisation, and use the system of Equation (\ref{eq:grid_corec_eq2}) to find the optimal gridding function $C(u)$ corresponding to this choice of $h$.

We perform the optimisation by representing $h(x)$ on a set of $N_g+1$ equally--spaced points from $x=0$ to $x=x_0$. From $x=0$ to $x=x_0$, $h(x)$ increases smoothly from $h(0)=1$. This function is also symmetric about $x=0$, so that it has a total of $(2N_g+1)$ points. For the variable $\nu$, we use a grid of $M_g$ equally--spaced points in the interior of the interval $0<\nu<0.5$. For $C(u)$, as a result of its symmetry, there are $2WM_g$ points. These grids need be of only modest size: $N_g=32$ and $M_g=16$ suffice for the examples presented here.

We use the following procedure to generate $E(h(x))$:
\begin{enumerate}
	\item From an initial $h(x)$ of $(N_g+1)$ values running from $x=0$ to $x=x_0$, we calculate $A_{r'r}$ from Equation (\ref{eq:All'}), using a trapezoidal rule to calculate the integral. We can then determine $A^{-1}_{r'r}$.
	\item We next evaluate $C_r(\nu)$ using Equation (\ref{eq:grid_corec_eq2}), again using a trapezoidal rule for the integral on the right-hand side.
	\item Next, calculate $\ell(x)$ from Equation (\ref{eq:l(x)}), approximating the integral over $\nu$ by a simple sum over the points at which $C_r(\nu)$ has been determined.
	\item Finally, we calculate $E$ from Equation (\ref{eq:Error_simple}), using the trapezoidal rule for the integral over $x$.
\end{enumerate}
Since we have essentially expressed $E$ in terms of the samples of $h(x)$ for $0 \leq x \leq x_0$ alone, we may use the Levenberg--Marquardt algorithm \citep{more1978levenberg} to perform the minimisation of $E$. By making use of modern \texttt{Python} libraries, and if we identify the $E$ function with the function \texttt{leastsq} from package \texttt{scipy.optimise}, it returns the optimised $h(x)$ with $(N_g+1)$ values.

Although the optimisation gives $h(x)$ sampled on a coarse grid between $0$ and $x_0$, we can still calculate $C(u)$ at arbitrary values of $u$ using Equation (\ref{eq:grid_corec_eq2}). The values of $h(x)$ and $\ell(x)$ can also be calculated at arbitrary points over the entire map $-0.5\leq x \leq 0.5$ by using Equations (\ref{eq:l(x)}) and (\ref{eq:h(x)}).

An initial guess for $h(x)$ is needed to begin the optimisation. We introduce the notation $h_W(x)$ to distinguish between correcting functions with differing support widths. The choice $h(x) = 1$ works well if $W \leq 4$. For $W > 4$ the minimisation does not converge satisfactorily. This problem can be overcome by providing the initial $h_{W}(x)$ as a function of the optimised $h_{(W-1)}(x)$ and $h_{(W-2)}(x)$:
\begin{enumerate}
	\item $h_{W}(x) = 1$ for $W\leq 4$;
	\item $h_{W}(x) = h_{(W-1)}^2(x)/h_{(W-2)}(x)$ for $W > 4$.
\end{enumerate}

\section{Calculation of least--misfit gridding functions for $W\geq 12$}\label{app:2}

As part of the optimisation algorithm it is necessary to compute the values of $C_r(\nu)$ from the current trial value of $h(x)$. The system in Equation (\ref{eq:grid_corec_eq}) involves the matrix $\mathbf{A}$, and its elements are given in Equation (\ref{eq:All'}). Since these elements are independent of $\nu$, the solutions for different values of $\nu$ can be found using a solver for linear systems, with the same coefficient matrix for differing right-hand sides. Unfortunately, for large values of $W$, the condition number of $\mathbf{A}$ increases rapidly, causing loss of precision and failure of the linear solver.

Equations (\ref{eq:grid_corec_eq}) can be regarded as the normal equations for solving the linear least--squares problem so as to minimise $E$ in Equation (\ref{eq:Error_simple}) for $C_r(\nu)\equiv C(r-\nu)$ when $h(x)$ is given. Instead of forming the normal equations, however, we can solve the least--squares problem directly. Upon substituting Equation (\ref{eq:l(x)}) into (\ref{eq:Error_simple}) and writing out the complex exponential in terms of its real and imaginary parts, we have
\begin{align}
	E &= \frac{1}{2x_0}\int_0^1\mathrm{d}\nu\int_{-x_0}^{x_0}\mathrm{d}x\\ \nonumber
	&\resizebox{0.5\textwidth}{!}{$\left[\left(1-h(x)\sum_r C_r(\nu)\cos[2\pi(r-\nu)x]\right )^2 + \left(h(x)\sum_r C_r(\nu)\sin[2\pi(r-\nu)x]\right)^2\right]. \nonumber$}
\end{align}
Since the integrand is non--negative at every value of $\nu$, we can consider each value separately and minimise
\begin{equation}
	\resizebox{0.5\textwidth}{!}{$\int_{-x_0}^{x_0}\mathrm{d}x\left[\left(1-h(x)\sum_r C_r(\nu))\cos[2\pi(r-\nu)x]\right )^2 + \left(h(x)\sum_r C_r(\nu)\sin[2\pi(r-\nu)x]\right)^2\right]$}. \nonumber
\end{equation}
Upon discretising the problem by approximating the integral by a sum over $M$ samples $x_i$ distributed over the interval of integration, we can minimise
\begin{equation}
	\resizebox{0.5\textwidth}{!}{$\sum_{i=1}^M\left[\left(1-h(x_i)\sum_r C_r(\nu))\cos[2\pi(r-\nu)x_i]\right )^2 + \left(h(x_i)\sum_r C_r(\nu)\sin[2\pi(r-\nu)x_i]\right)^2\right]$}. \nonumber
\end{equation}
This may be written as $||\mathbf{d}-\mathbf{Bc}||^2$ where $c_r=C_r(\nu)$, $d_i = \begin{cases}
      1 & \text{if $1\leq i \leq M$}\\
      0 & \text{if $M+1\leq i \leq 2M$}
   \end{cases}$ and
\begin{equation}\label{eq:B_ir}
B_{ir} =
  \begin{cases}
         h(x_i)\cos[2\pi(r-\nu)x_i] & \text{if $1\leq i \leq M$}\\
	h(x_i)\sin[2\pi(r-\nu)x_i] & \text{if $M+1\leq i \leq 2M$}  \end{cases}	
\end{equation}
For each value of $\nu$, the linear--least squares problem for $\textbf{c}$ given $\textbf{B}$ and $\textbf{d}$ can be solved by factorizing $\mathbf{B}$ using the QR algorithm (as is done by the \texttt{dgels()} routine in \texttt{LAPACK}). This procedure does not require the matrix $\mathbf{A}$ in Equation (\ref{eq:grid_corec_eq}), which is numerically badly conditioned; consequently the calculation can be done using double precision arithmetic for $W\leq 14$.

Values of $W$ that are even higher may be considered using an extended precision package such as \texttt{gmpy2} in \texttt{Python}. It was convenient to reformulate the optimization as an eigenvalue problem in this case, and for the least--misfit gridding function with $W=20$ and $x_0=0.25$ the integrated map error was $6.62 \times 10^{-43}$.

\section{Least--misfit gridding function with small values of $W$}\label{app:3}

For $W=1$, each visibility point is gridded onto its nearest neighbour. Figure (\ref{fig:W15demo}) shows that the least--misfit gridding function $C(u)$ in this case is approximately equal to one over the range $|u|<0.5$. The fact that it is not exactly one implies that, when the visibility is exactly on a gridpoint, a higher weight is preferable than when the visibility is midway between grid points. Since $C(u)=0$ for $|u|>0.5$, the gridding function is discontinuous at $|u|=0.5$.

For $W=2$, each visibility point is gridded onto its two nearest neighbours. The least--misfit function in this case differs slightly from the simple linear interpolation function $C(u)=1-|u|$ for $|u|\leq 1$. The orange curve in the middle left panel of Figure (\ref{fig:W15demo}) is not a straight line, and does not pass through one when $u=0$ and zero when $u=1$. This tells us that the gridding function is discontinuous at $|u|=1$, since its value is zero for all $|u|\geq 1$. Since $C$ is even, it too has a discontinuity in its slope at $u=0$. When this gridding function is used in conjunction with the correction function, the integrated map error from zero to $x_0=0.25$ is smaller than for conventional linear interpolation; see Figure (\ref{fig:many_maperr_demo}).

Based on these two cases, we expect that $C(u)$ (and its derivatives) may be discontinuous at integer values of $u$ when $W$ is even, and may be discontinuous at half-integer values of $u$ (i.e., values half-way between the integers) when $W$ is odd.

For $W=3$, visible discontinuities remain in the least--misfit gridding function $C(u)$ at $|u|=0.5$ and $|u| = 1.5$. This behaviour is a natural extension of the situation for $W=1$, and may also be understood by considering the set of three grid points onto which each visibility is distributed. According to which side of the midpoint between two grid points a baseline falls, the visibility is gridded onto a different set of three points. For example, if the baseline is $\lfloor u_0 \rfloor+0.49$ (where $\lfloor u_0 \rfloor$ is an integer), the visibility is distributed among $\lfloor u_0 \rfloor-1$, $\lfloor u_0 \rfloor$ and $\lfloor u_0 \rfloor+1$, whereas if the baseline is $\lfloor u_0 \rfloor+0.51$, the visibility is distributed among $\lfloor u_0 \rfloor$, $\lfloor u_0 \rfloor+1$ and $\lfloor u_0 \rfloor+2$. Since the value of the gridding function at the edge of its support, i.e., $C(\pm 1.5)$, is non-negligible, the way in which the visibility is distributed among the grid points is expected to change discontinuously as the baseline passes through $\lfloor u_0 \rfloor+0.5$; this is because one grid point disappears from the set and is replaced by another.

For $W\geq 4$, these discontinuities in value and derivatives cease to be visible. In particular, the values of $C(u)$ at the edge of the support $|u|=W/2$ are very small.

\section{34-source simulated data}\label{app:4}
Table (\ref{tab:testcard_table}) shows the location and flux of the 34 simulated point sources used in the numerical experiments. Sources are scattered about the phase centre with different fluxes across the full field of view.

\begin{table}
\centering
 \caption{Locations and fluxes of 34 simulated point sources. $X$ and $Y$ are specified in pixel numbers, and represent the distances from the image centre $(0,0)$ to the corresponding point sources.}
 \label{tab:testcard_table}
 \resizebox{\columnwidth}{!}{%
 \begin{tabular}{l|ccc|l|ccc}
  \hline
   Index & X & Y & Flux (Jy) & Index & X & Y & Flux (Jy)\\
  \hline
  1 & 0 & 0 & 2 & 18 & 0 & 270 & 1 \\ \hline
  2 & 0 & 15 & 2 & 19 & 0 & 330 & 1 \\ \hline
  3 & -120 & 180 & 2 & 20 & 330 & 0 & 1 \\ \hline
  4 & 150 & -150 & 2 & 21 & 0 & -330 & 1 \\ \hline
  5 & 300 & 90 & 2 & 22 & -330 & 0 & 1 \\ \hline
  6 & -90 & 300 & 2 & 23 & 270 & 270 & 1 \\ \hline
  7 & 90 & -90 & 1 & 24 & 270 & -270 & 1 \\ \hline
  8 & -90 & 90 & 1& 25 & -270 & 270 & 1 \\ \hline
  9 & -90 & -90 & 1 & 26 & -270 & -270 & 1 \\ \hline
  10 & 180 & 90 & 1 & 27 & 390 & 390 & 3\\ \hline
  11 & 180 & 180 & 1 & 28 & 390 & -390 & 3 \\ \hline
  12 & 180 & -180 & 1 & 29 & -390 & -390 & 3 \\ \hline
  13 & -180 & 180 & 1 & 30 & -390 & 390 & 3 \\ \hline
  14 & -180 & -180 & 1 & 31 & 345 & 0 & 2 \\ \hline
  15 & 270 & 0 & 1 & 32 & -345 & 0 & 2\\ \hline
  16 & 0 & -270 & 1 & 33 & 0 & -345 & 2 \\ \hline
  17 & -270 & 0 & 1 & 34 & 0 & 345 & 2 \\ \hline
  \hline
 \end{tabular}
 }
\end{table}

Sources 27, 28, 29 and 30 are at four corners of the field, and are chosen to have the largest fluxes of the 34 sources. 

\section{Generation of the Look--up table}\label{app:5}

The look--up table for the least--misfit gridding function is assembled as follows:
\begin{itemize}
	\item Create $\nu \in [0,1)$ as a set of $M_s$ numbers spaced equally between $0$ and $1$. We refer to $M_s$ as the `sampling rate' in this paper. 
	\item If $W\leq 11$ then, for each value of $\nu$, calculate a set of $W$ gridding function values according to Equation (\ref{eq:grid_corec_eq2}) using matrix $\textbf{A}$, which can be used repeatedly as it is not dependent on $\nu$.
	\item If $W> 11$ then, for each value of $\nu$, calculate a set of $W$ gridding function values according to Equation (\ref{eq:B_ir}) using matrix $\textbf{B}$. The $\textbf{B}$ matrix must be recalculated for each $\nu$, increasing the computational cost.
\end{itemize}

\section{Effect of the Look--up table}\label{app:6}

Figure (\ref{fig:Image_misfit_leastmisfit_fullmap_sqaure_testcard}) shows that, if the gridding function is calculated precisely, the RMS misfit between the DFT and the FFT is proportional to the square root of the map error $\ell(x)$, as expected from equation (\ref{eq:ell}). When, however, we store the gridding function $C(u)$ for $-W/2\leq u < W/2$ in a table with $W M_s$ points over this range, and use nearest-neighbour lookup to perform the convolution as shown in Figure (\ref{fig:Leastmisfit_W7_nearest}), the performance is degraded unless a large value of $M_s \gtrsim 10^6$ is used. The misfit is also no longer constant over the map, but tends to increase away from the centre of the map as a result of performing the lookup.

\begin{figure*}
\begin{center}
\includegraphics[width=0.4\textwidth]{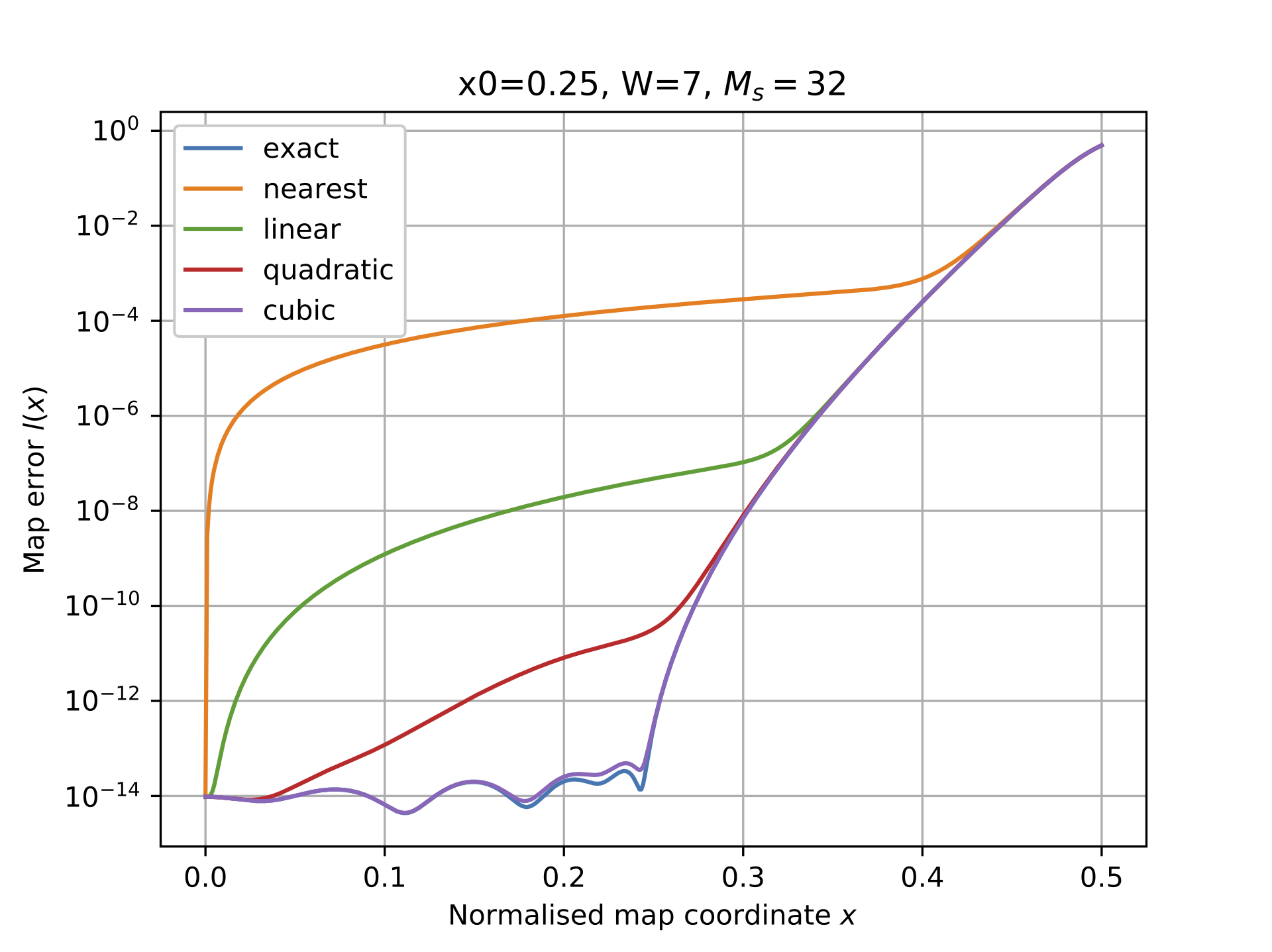}
\includegraphics[width=0.4\textwidth]{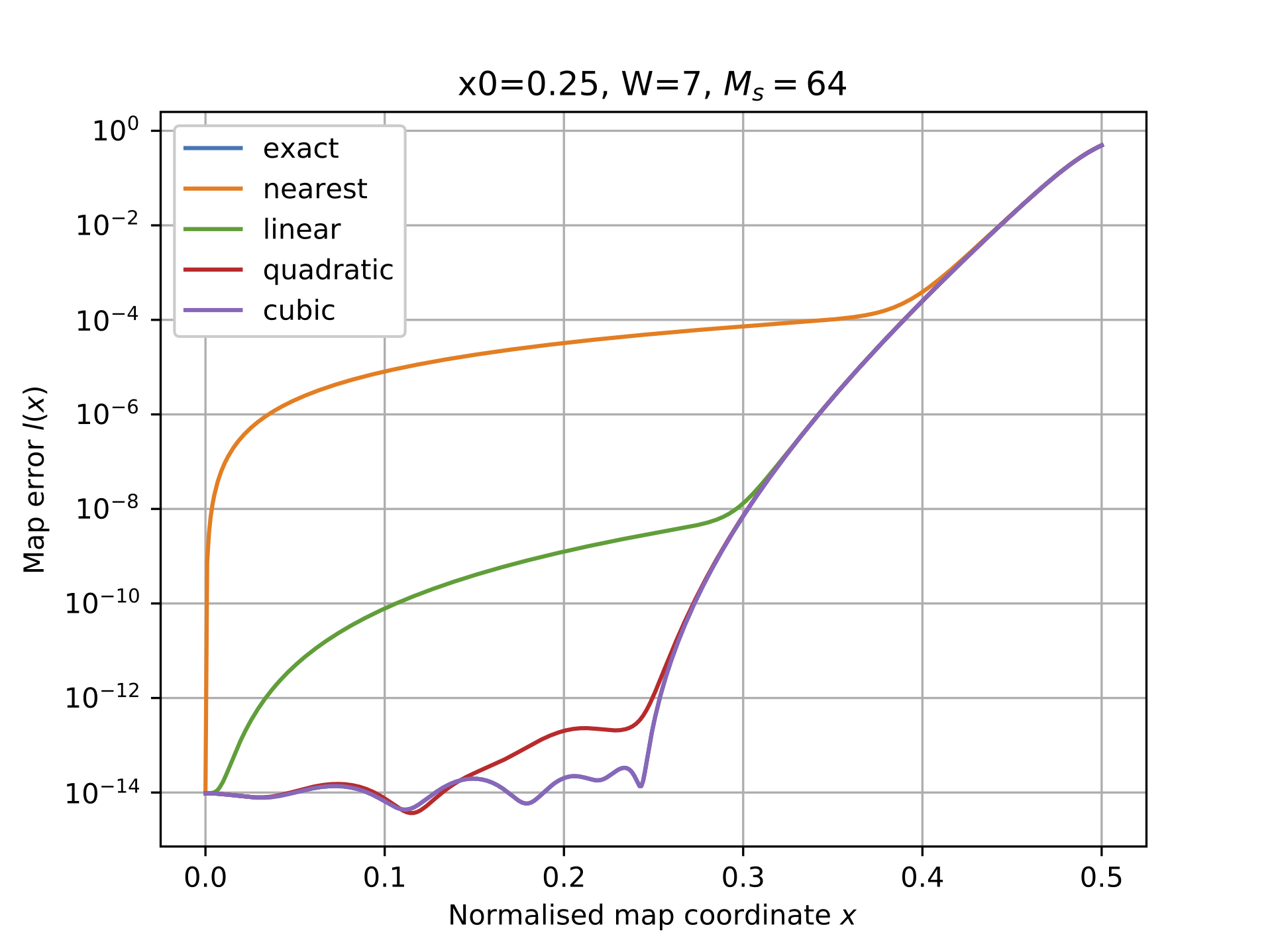}
\includegraphics[width=0.4\textwidth]{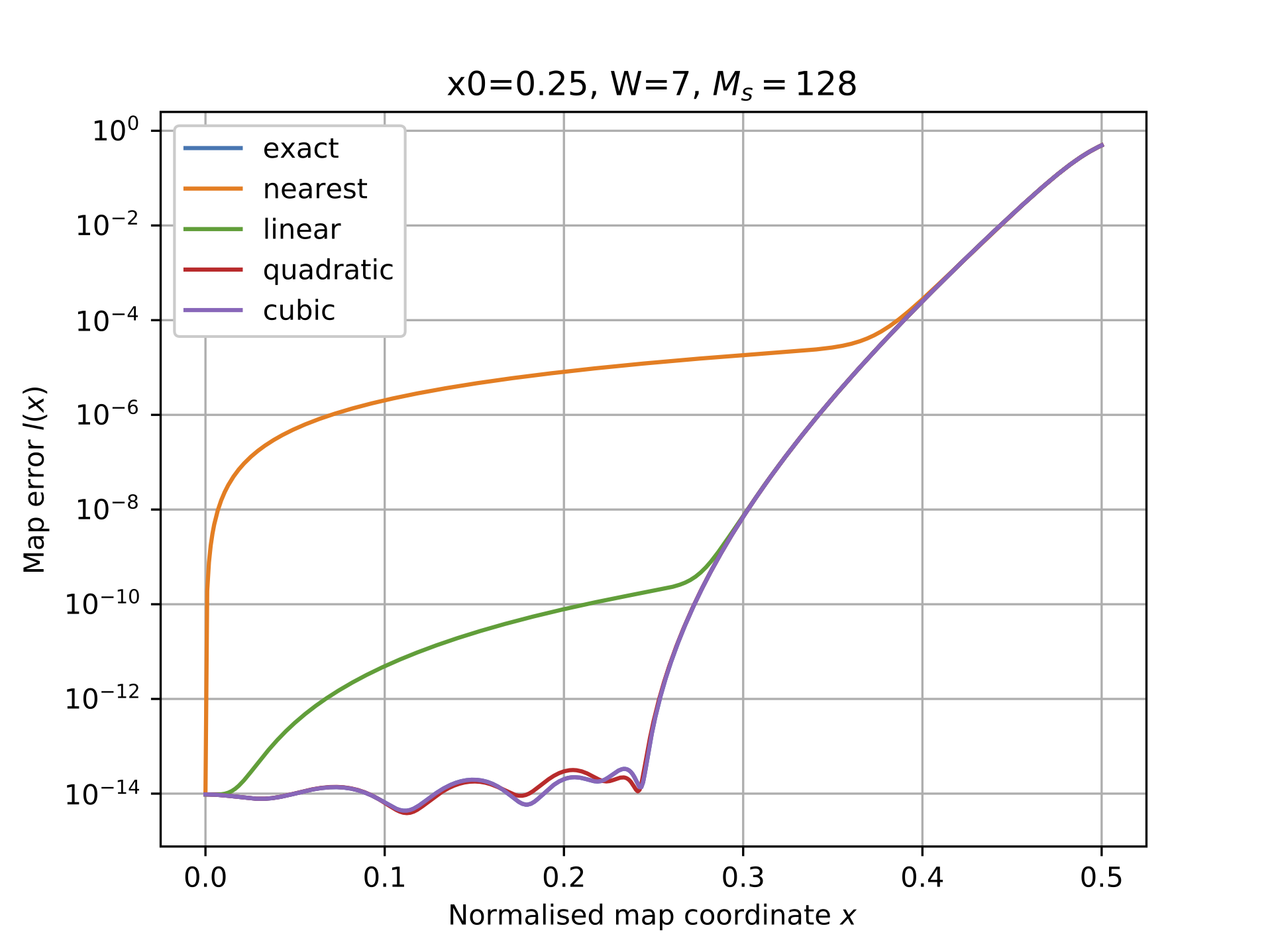}
\includegraphics[width=0.4\textwidth]{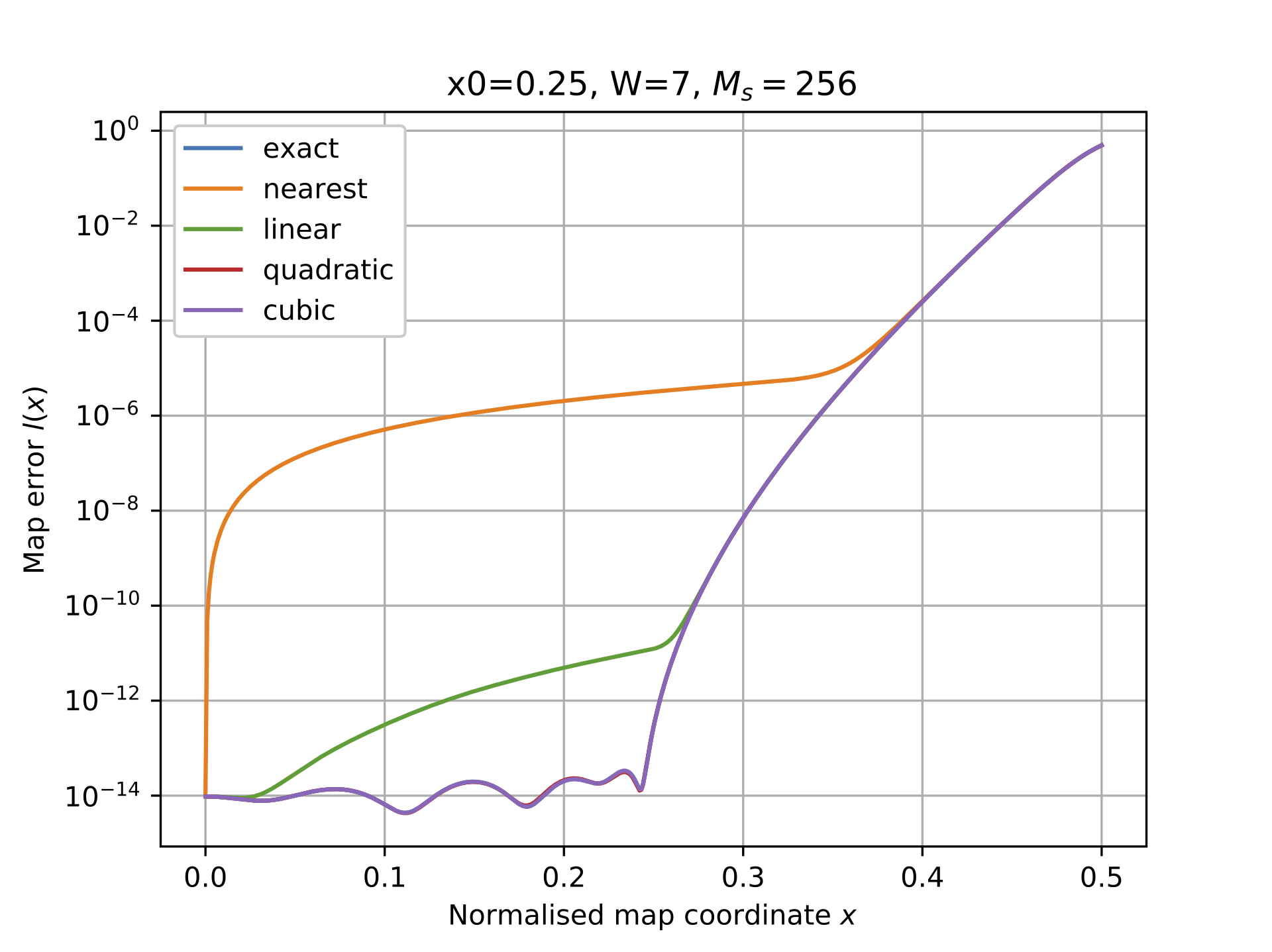}
\caption{Map error $\ell(x)$ for different interpolation schemes and lookup table sampling rate $M_s$ using the least-misfit gridding function for $W=7$ and $x_0=0.25$. }
	\label{fig:map-error-interp}
\end{center}
\end{figure*}

In this appendix we account for these effects, and investigate how the degradation depends on the value of $M_s$. As well as nearest-neighbour lookup, the higher-order interpolation schemes for calculating $C(u)$ from its sampled values are also of interest.

The key observation is that equation (\ref{eq:l(x)}) for the map error $\ell(x)$ holds for any choice of $h(x)$ and $C_r(\nu)\equiv C(r-\nu)$. If these are set equal to the optimal $h^{\mathrm{(opt)}}(x)$ and $C_r^{\mathrm{(opt)}}(\nu)$ for the least--misfit functions, we find results such as those in Figure (\ref{fig:W14demo}). When table lookup or interpolation are used, we are effectively using a different gridding function $C_r(\nu)$ which coincides with the optimal function at the tabulated points, but which takes different values away from these points. By substituting the interpolated function for $C_r(\nu)$ and the optimal correction $h^{\mathrm{(opt)}}(x)$ into (\ref{eq:l(x)}), it is possible to evaluate $\ell(x)$ for the various schemes. We can also use equation (\ref{eq:Error_simple}) to calculate the mean square misfit over the map region $-x_0\leq x \leq x_0$; its square root provides an RMS error bound.

We consider the following interpolation methods, using a table of $C(\nu_k)$ where $\nu_k=k/M_s$ and $k$ takes integer values lying between $-WM_s/2$ and $WM_s/2$.
\begin{itemize}
	\item Nearest neighbour table lookup, in which $C(\nu)$ is approximated by $C(\nu_k)$ where $\nu_k$ is the entry in the table closest to $\nu$. This is a piecewise polynomial approximation of degree $d=0$,
	\item Linear interpolation, of degree $d=1$
	\item Quadratic interpolation, of degree $d=2$
	\item Cubic interpolation, of degree $d=3$
\end{itemize}
For $d\geq 1$ we use a polynomial of degree $d$ which passes through $\nu_k$, $\nu_{k+1}$,..., $\nu_{k+d}$ evaluated at $\nu$, where $\nu_k\leq \nu < \nu_{k+1}$. The computation can be performed efficiently using the standard Newton interpolation formula (see Abramowitz and Stegun 25.2.28 and 25.2.29). If the function that is to be approximated has continuous derivatives of order up to $d+1$, the error in the approximation is $\mathcal{O}(h^{d+1})$, where $h\equiv M_s^{-1}$ is the separation between the tabulated points.

Figure \ref{fig:map-error-interp} shows how the map error function $\ell(x)$ varies for these interpolation schemes, as well as for the exact gridding function, in the case of the least-misfit gridding function for $W=7$ and $x_0=0.25$. Interpolation schemes generally increase the map error relative to the exact gridding function over almost all of the map, with the degradation getting worse away from the map centre. Re-plotting of the figures on log-log axes reveals that, when the error due to interpolation is greater than that for the exact gridding function, $\ell(x)\sim x^{2(d+1)}$. Thus the RMS difference between the DFT and FFT maps in these regions behaves as $x^{(d+1)}$.

\begin{figure}
\centering\includegraphics[width=\columnwidth]{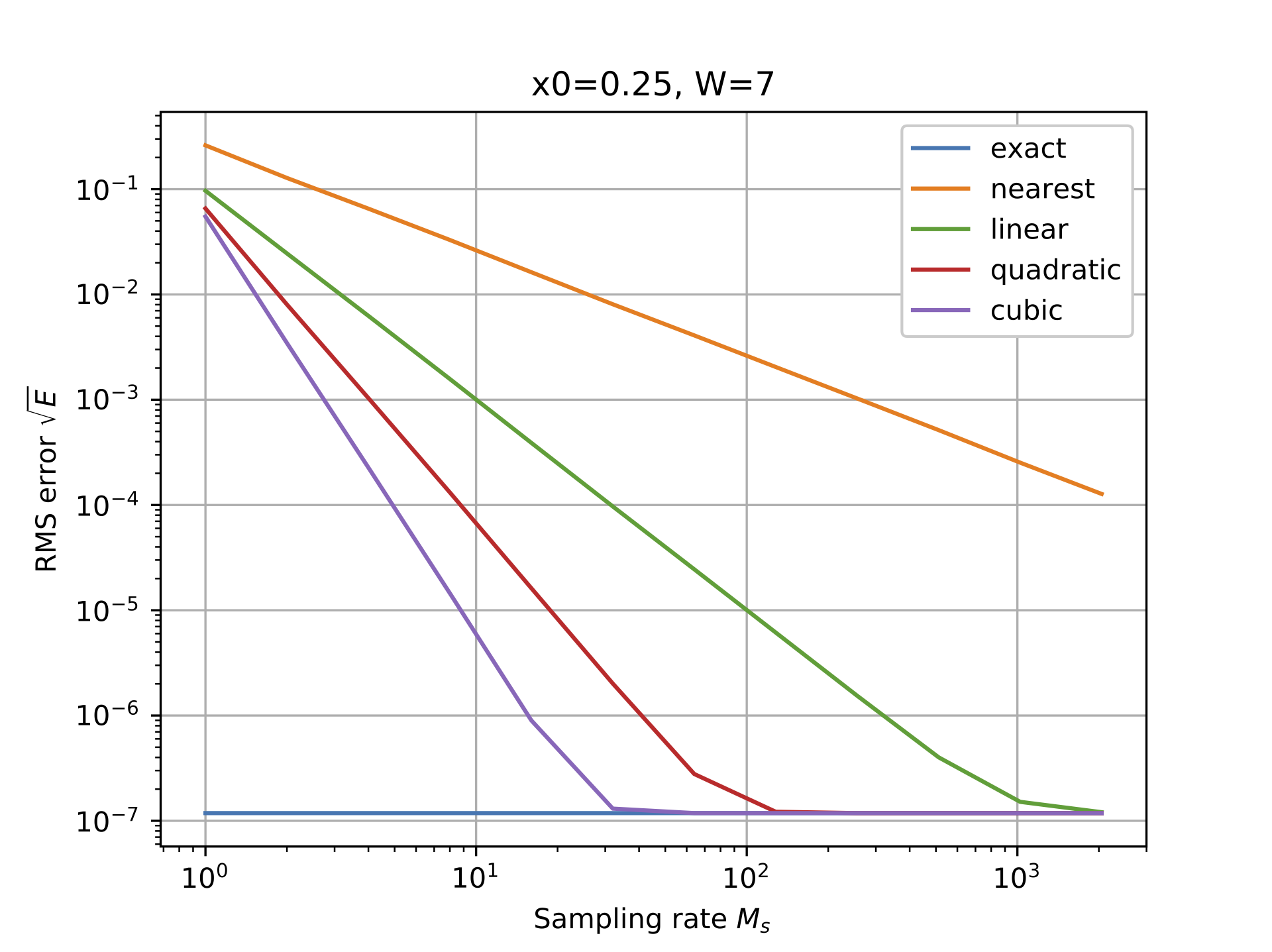}
	\caption{Root mean square error $\sqrt{E}$ over the portion of map retained in various interpolation schemes for the least-misfit gridding function with $W=7$ and $x_0=0.25$.}
	\label{fig:rms-interp-err}
\end{figure}

We see also from the graphs that, as $M_s$ increases, the rate at which $\ell(x)$ approaches the exact gridding function depends strongly on the choice of $d$. A plot of $\sqrt{E}$ where $E$ is given by equation (\ref{eq:Error_simple}), provides a bound on the RMS error over $-x_0\leq x < x_0$. This is shown in Figure \ref{fig:rms-interp-err}. Interpolation of degree $d$ leads to an RMS error dependent on $M_s^{-(d+1)}$ above the minimum provided by the exact gridding function. Based on such graphs, we may choose the size of the table so that the results of the interpolation are indistinguishable from results using the exact gridding function. In the example considered, $M_s\approx 10^6$ for nearest-neighbor interpolation, and $10^3$ for linear interpolation.

The choice of interpolation scheme depends on the trade-off between the time required to fetch data from a lookup table of a given size, and the time to do the interpolation of a given degree. If the computation time dominates, a low-degree interpolation and a large lookup table is preferable. If memory access time dominates, the ability to fit a smaller table into high speed cache may favour interpolation of higher degree.



\label{lastpage}
\end{document}